%% file: main.tex
\documentclass[10pt,journal,compsoc]{IEEEtran}

\usepackage{booktabs}
\usepackage{graphicx}
\usepackage[table,xcdraw]{xcolor}
\usepackage{dblfloatfix}
\newcommand{\todo}[1]{}
\renewcommand{\todo}[1]{{\color{red} TODO: {#1}}}
\usepackage{comment}
\usepackage{amssymb}
\usepackage{multirow}
\usepackage{lscape}

\ifCLASSOPTIONcompsoc
  \usepackage[nocompress]{cite}
\else
  \usepackage{cite}
\fi
\ifCLASSINFOpdf
\else
\fi

\begin{document}
%

\title{A Multi-dimensional Study of Requirements Changes in Agile Software Development Projects}

%
%
%
%

\author{Kashumi Madampe,~\IEEEmembership{Graduate Student Member,~IEEE,}
        Rashina~Hoda,~\IEEEmembership{Member,~IEEE,}
        and~John~Grundy,~\IEEEmembership{Senior Member,~IEEE}
\IEEEcompsocitemizethanks{\IEEEcompsocthanksitem K. Madampe, R. Hoda, and J. Grundy are with the HumaniSE Lab at Department
of Software Systems and Cybersecurity, Faculty of Information Technology, Monash University, Wellington Road, Clayton, VIC 3800, Australia.\protect\\
E-mail: kashumi.madampe@monash.edu}
\thanks{Manuscript received December 4, 2020; revised Month Date, 2020.}}

%
%

\markboth{Submitted to IEEE Transactions on Software Engineering}%
{Shell \MakeLowercase{\textit{et al.}}: Bare Advanced Demo of IEEEtran.cls for IEEE Computer Society Journals}
%



\IEEEtitleabstractindextext{%
\begin{abstract}

Agile processes are now widely practiced by software engineering (SE) teams, and the agile manifesto claims that agile methods support responding to changes well. However, no study appears to have researched whether this is accurate in reality. Requirements changes (RCs) are inevitable in any software development environment, and we wanted to acquire a holistic picture of how RCs occur and are handled in agile SE teams in practice. We also wanted to know whether responding to changes is the only or a main reason for software teams to use agile in their projects. To do this we conducted a mixed-methods research study which comprised of interviews of 10 agile practitioners from New Zealand and Australia, a literature review, and an in-depth survey with the participation of 40 agile practitioners world-wide. Through this study we identified different types of RCs, their origination including reasons for origination, forms, sources, carriers, and events at which they originate, challenging nature, and finally whether agile helps to respond to changes or not. We also found that agile teams seem to be reluctant to accept RCs, and therefore, they use several mitigation strategies. Additionally, as they accept the RCs, they use a variety of techniques to handle them. Furthermore, we found that agile allowing better response to RCs is only a minor reason for practicing agile. Several more important reasons included being able to deliver the product in a shorter period and increasing team productivity. Practitioners stated this improves the agile team environment and thus are the real motivators for teams to practice agile. Finally, we provide a set of practical recommendations that can be used to better handle RCs effectively in agile software development environments.

\end{abstract}

}

\maketitle

\IEEEdisplaynontitleabstractindextext

%
\IEEEpeerreviewmaketitle


\ifCLASSOPTIONcompsoc
\IEEEraisesectionheading{\section{Introduction}\label{sec:Intro}}
\input{sections/introduction}

\else
\fi

\label{sec:motivation}

\section{Research Questions}

\input{sections/RQs}
\label{sec:RQs}

\input{sections/method}
\label{sec:Mtd}

\section{Results}
\label{sec:Results}

    \subsection{What  are  the  Types  of  Requirements  Changes Agile Teams Face? (RQ1)}
    \label{sec:type}
    \input{sections/type}
    
        \subsection{Why are Requirements Changes Needed? (RQ2.1)}
        \label{sec:reason}
        \input{sections/reason}
        
        \subsection{What are the Sources of Requirements Changes? [Artefacts] (RQ2.2) }
        \label{sec:source}

\input{sections/source}

        \subsection{Where do Requirements Changes Originate? [Events] (RQ2.3)}
        \label{sec:where}
        \input{sections/where}

        \subsection{Who brings Requirements Changes to the Team? [Carriers] (RQ2.4)}
        \label{sec:who}
        \input{sections/who}

        \subsection{In Which Forms are Requirements Changes Documented? (RQ2.5)}
        \label{sec:form}

\input{sections/form}

    \subsection{How Challenging are the Requirements Changes? (RQ3)}
    \label{sec:challenge}
    \input{sections/challenge}

    \subsection{Does Agile Help in Responding to Requirements Changes? (RQ4)}
    \label{sec:agile_helps}
    \input{sections/agilehelps}

\section{Discussion}
\label{sec:discussion}
    \input{sections/discussion}

\section{Recommendations}
    \label{sec:recommendations}
    \input{sections/recommendations}

\section{Related Work}
\label{sec:RW}
\input{sections/relatedwork}

\section{Conclusion}
\label{sec:conclusion}
\input{sections/conclusion}

\section*{Acknowledgement}
This work is supported by a Monash Faculty of IT scholarship. Grundy is supported by ARC Laureate Fellowship FL190100035. Also, our sincere gratitude goes to \emph{Agile Alliance} and all the participants who took part in this study.

\bibliographystyle{ieeetr}
\bibliography{main}


%

\begin{IEEEbiography}[{\includegraphics[width=1in,height=1.25in,clip,keepaspectratio]{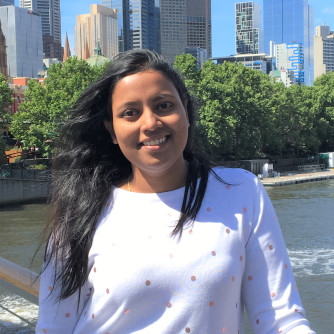}}]{Kashumi Madampe}is a Graduate Student Member of IEEE and is with Monash University, Melbourne, Australia. Ms. Madampe did part of her current research at The University of Auckland, New Zealand. Prior to the PhD candidature, she was in the software development industry as a project manager and a business analyst. Her research interests are requirements engineering, human and social aspects of software engineering, software repository mining, grounded theory, and natural language processing. She serves as the ASE2021 publicity and social media chair and CHASE2021 social media chair. More details about her research can be found at https://kashumim.com. Contact her at kashumi.madampe@monash.edu.
\end{IEEEbiography}

\begin{IEEEbiography}[{\includegraphics[width=1in,height=1.25in,clip,keepaspectratio]{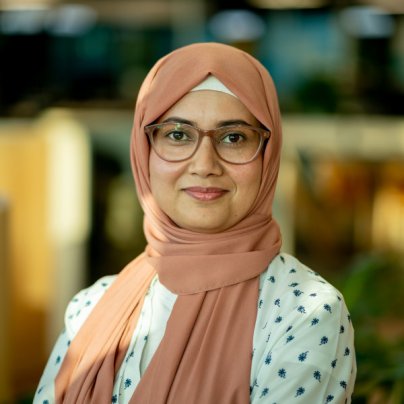}}]{Rashina Hoda}is an associate professor in software engineering at Monash University, Melbourne, Australia. She received her PhD in computer science from Victoria University of Wellington, New Zealand. Her areas of research interests include human and social aspects of software engineering, grounded theory, and serious game design. She serves on the IEEE Transactions on Software Engineering review board, IEEE Software Advisory Board, as ICSE2021 social media co--chair, CHASE 2021 program co--chair, and XP2020 program co--chair. More details about her research can be found at https://rashina.com. Contact her at rashina.hoda@monash.edu.

\end{IEEEbiography}

\begin{IEEEbiography}[{\includegraphics[width=1in,height=1.25in,clip,keepaspectratio]{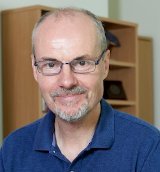}}]{John Grundy}received the BSc (Hons), MSc, and PhD degrees in computer science from the University of Auckland, New Zealand. He is an Australian Laureate fellow and a professor of software engineering at Monash University, Melbourne, Australia. He is an associate editor of the IEEE Transactions on Software Engineering, the Automated Software Engineering Journal, and IEEE Software. His current interests include domain--specific visual languages, model--driven engineering, large-scale systems engineering, and software engineering education. More details about his research can be found at https://sites.google.com/site/johncgrundy/. Contact him at john.grundy@monash.edu.
\end{IEEEbiography}




\end{document}

%% file: sections/introduction.tex

\IEEEPARstart{R}{equirements} changes are inevitable in software engineering. However, requirements changes (RCs) pose a risk to the cost, quality, and schedule of a  project \cite{McGee2009ATaxonomy}. Agile methods claim to help to enhance the competitive advantage of the customer by responding to and accepting RCs even late in development \cite{Beck2001ManifestoDevelopment,Hoda2018TheDevelopment}. Indeed, the top three metrics stated by the \emph{14th State of Agile} -- the largest annual industry survey report -- are when \emph{customer satisfaction, business value}, and \emph{on-time delivery} are fulfilled \cite{202014thAgile}. 





Imagine an agile software development team receiving different types of RCs, from multiple sources by numerous stakeholders, in various forms, and during agile ceremonies/events. How challenging might this be to the team? Are the teams using agile methods because of its ability to respond to RCs? Do agile methods actually help the team to respond to RCs in reality? A preliminary study of 10 interviews with agile practitioners in New Zealand and Australia, revealed such scenarios motivating us to conduct further research in order to construct a holistic picture of RCs in agile software development, the challenging nature of receiving them, better understand how teams receive and handle them in practice, and to provide a set of recommendations on better handling RCs during agile software development. 

To the best of our knowledge, none of the existing studies of agile software development have comprehensively investigated these aspects. Prior studies 
\cite{McGee2011SoftwareStudy,McGee2009ATaxonomy} resulted in an RC taxonomy for software development based on origination sources categorized as RCs originating from \textit{market, organization, project vision, specification,} and \textit{solution}. Nurmuliani et al. \cite{Nurmuliani2004AnalysisCycle} presents types of RCs, reasons why they originate, and sources from which RCs originate as their key findings. They categorized types of RCs as additions, deletions, and modifications of requirements. Inpirom and Prompoon classified RCs according to analysis and design of software artefacts \cite{Inpirom2013DiagramUML}. But these weren't applied to agile methods and handling RCs during agile software development.  Saher et al.  \cite{Saher2018} describe RCs in terms of \textit{time of change, type, reason,} and \textit{origin} of RC. However only one study they used actually focused on agile methods.

In order to answer these outstanding questions about handling RCs in agile software development, we need a holistic understanding of how RCs originate, are handled and are responded to in agile environments. To do this, we conducted an in-depth survey\footnote{Approved by Monash Human Research Ethics Committee. Approval Number: 23578} study with the participation of 40 agile practitioners across Asia (N=26), Oceania (N=9), North America (N=3), and Europe (N=2). 
Key findings include that the majority of RCs received are functional; more RCs are human-centric than software-centric; RCs originate mostly during daily standups; software-centric and human-centric RCs need to be handled in different ways; a number of issues affect teams such as stress when receiving RCs, reluctance to receive new RCs, and a need for well-defined acceptance criteria to handle RCs  used by agile teams; and the customer produces the RCs most but not all of the time. Based on these findings, we identified a number of recommendations that could improve the handling of RCs by agile software development teams.


Therefore, the key contributions of this work include:
\begin{itemize}
 \item Insights from 10 interviews and in-depth survey of 40 experienced agile practitioners to understand how they handle RCs in real software projects;
 \item A set of metrics to measure the challenging nature of RCs in agile projects;
    \item A taxonomy of RCs in agile that we present as a conceptual framework in terms of human and non-human aspects of RCs resulting from descriptive analysis. These include reasons, sources, events where the RCs occur, types and forms, carriers, challenging metrics of RCs;
    \item Whether agile truly enables better responding to RCs or not; 
    \item A set of practical recommendations for agile teams to better handle diverse RCs

\end{itemize}

The rest of this paper is structured as follows. Section \ref{sec:RQs} provides the research questions we used to guide this research. Section \ref{sec:SD} provides the research method including the survey design, data collection, and analysis. The answers to the research questions are given in Section \ref{sec:Results} followed by the discussion and threats to validity in Section \ref{sec:discussion}. Section \ref{sec:recommendations} provide recommendations for agile practitioners and future work of our study. Related work is given in Section \ref{sec:RW} followed by the conclusion in Section \ref{sec:conclusion} providing a summary of our findings.


%% file: sections/RQs.tex
\begin{figure*}[]
    \centering
    \includegraphics[width=\textwidth]{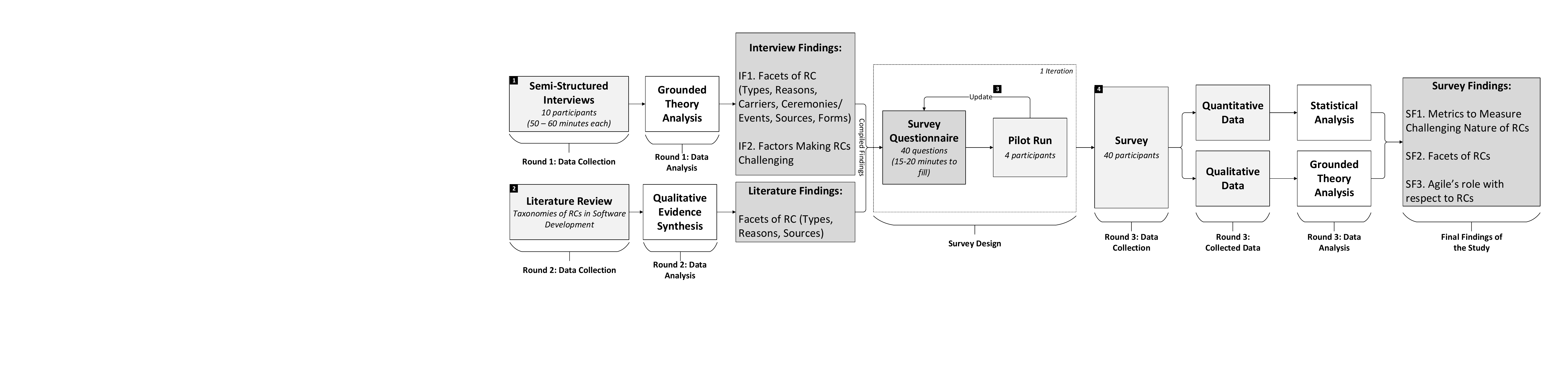}
    \caption{Mixed-Methods Research Including Semi-Structured Interviews, Literature Review, and In-Depth Survey (IF: Interview Finding; SF: Survey Finding)}
    \label{fig:method}
\end{figure*}


We wanted to gain a comprehensive understanding of RCs in agile and to understand whether agile helps to respond to RCs. We derived the following research questions to guide our study with the help from our interview-based preliminary study (Section \ref{sec:interview}) findings and existing literature.

    \textbf{RQ1. What are the types of requirements changes agile teams face?}
    Several types of RCs were found from our preliminary study data analysis. We wanted understand these more comprehensively by using a larger sample of real world agile software development teams.
    
    \textbf{RQ2. Origin of requirements changes}
    We wanted to know why, what, when, and who are involved in producing RCs in an agile software project. We formulated the following sub-RQs:
\begin{itemize}
   
        \item \textit{RQ2.1. Why are RCs needed?}
    
        \item \textit{RQ2.2. What are the sources and frequency of RCs?}
        
        \item \textit{RQ2.3. Where do the RCs originate?}
        Apart from typical agile ceremonies, our preliminary study proposed that RCs originate after releasing the complete product as well. We wondered about other times and reasons where RCs originate as well as the frequency of the origin at the particular ceremony/event.
        
        \item \textit{RQ2.4. Who brings RCs and how often?}
        Not only the customer, but also other stakeholders bring RCs to the team according to our preliminary findings. We used this question to confirm the carriers we found and how often they bring RCs to the team.
        
        \item \textit{RQ2.5. In which forms are requirements changes documented?}
        Given that various ways exist to document requirements changes, we wanted to know the forms that agile teams use to document RCs.
    \end{itemize}
    
     \textbf{RQ3. How challenging are the requirements changes to handle?}
    Requirements changes  can be very challenging for all software development teams. We wanted to understand how challenging are RCs to teams using a set of metrics we defined to measure the nature of these challenges.
    
    \textbf{RQ4. Does Agile help to better respond to requirements changes?}
    As agile software development is supposedly popular because of its ability to allow a team to better respond to RCs, we wanted to know in what ways agile actually helps the teams  when handling RCs.
    

%% file: sections/method.tex
\section{Study Design}
\label{sec:SD}

The mixed-methods approach used for our study is outlined in Fig. \ref{fig:method}. This study was conducted with the participation of agile practitioners across the world. First we conducted an interview-based preliminary study to understand the landscape of RCs in agile projects. Findings of the preliminary interviews were combined with a literature review to construct the survey questionnaire. Once the questionnaire was finalized, a pilot run was conducted before distributing the survey to the agile community in the software development industry world-wide.

\subsection{Interview-Based Preliminary Study and Literature Review}\label{sec:interview}
We conducted 10 semi-structured (N=7 face-to-face, N=3 online) interviews with agile practitioners in New Zealand (N=8) and Australia (N=2). Each interview lasted 50-60 minutes. The participants had 19.75 years of mean total experience in software development  (min. total experience=2 years; max. total experience=56 years) and 7.26 years of mean total agile experience (min. agile experience=1 year; max. agile experience=18 years). Their job roles were Scrum Master (N=5), Business Analyst (N=2), Manager (N=2), Tester (N=1), Architect (N=1), Senior Consultant (N=1) and Head of Global Projects (N=1). Some participants played more than one role. The participants had experience in Scrum (N=10), XP(N=5), Scrum XP combo (N=4), Kanban (N=8), Feature Driven Development (N=3), Spotify (N=2), Dynamic Systems Development (N=1), and Company based Method (N=1).

We analysed the collected data from the interviews using Grounded Theory (GT) analysis procedures including open coding and constant comparison \cite{Strauss1998BasicsTechniques.}. The findings resulted in categories: \emph{reasons, types, sources, carriers}, and \emph{ceremonies/events}. 

We then conducted a literature review and combined findings from the interview-based study with closely related work of McGee and Greer's \cite{McGee2009ATaxonomy} and Nurmuliani et al.'s \cite{Nurmuliani2004AnalysisCycle} (summarized in Section \ref{sec:RW}) to develop an in-depth survey questionnaire. The survey consisted of 3 open-ended and 37 closed-ended questions. Furthermore, each close-ended question included one open-ended ``Other'' option as well for the participants to enter any other responses they wanted to include apart from the options we provided for the particular question. We followed Kitchenham et al.'s \cite{Kitchenham2002PreliminaryEngineering, Kitchenham2008PersonalSurveys} and Punter et al.'s \cite{Punter2003ConductingEngineering} guidelines to design the survey. We followed Smith et al.'s work \cite{Smith2013ImprovingSurveys} on ``improving developer participation rates in surveys'' to support survey distribution. Table \ref{tab:survey_questions} shows sample questions for each type. 

\begin{table*}[!t]
\centering
\caption{Sample Survey Questions}
\label{tab:survey_questions}
\begin{tabular}{@{}ll@{}}
\toprule
\multicolumn{1}{c}{\textbf{Question}}                                                                                                                                                                                                                                                                                                                                                                                                                                                                                                                                                                                                                                                                                              & \textbf{Question Type    }                                                                                 \\ \midrule
\begin{tabular}[c]{@{}l@{}}A requirements change can occur due to,\\ $<$Bug (Error arising in the codebase), Design improvement, Erroneous requirements, Functionality enhancement, \\ Missing requirements, Need for refactoring, Obsolete functionality, Outstanding technical debt, Product strategy, \\ Redundant functionality, Requirements clarification, Resolving conflicts, Scope reduction, Other$>$\end{tabular}                                                                                                                                                                                                                                                                                             & \begin{tabular}[c]{@{}l@{}}Close-ended (Multiple \\ selection with one\\ text entry)\end{tabular} \\ \\

Thinking about the specific (current/past) project, typically requirements changes make up to what percentage of \\ the total work items? \\
$<$0\% (We don't receive any requirements changes), Less than 25\%, 26-50\%, 51-75\%, More than 75\%$>$ & Close-ended (Single selection) \\ \midrule
\begin{tabular}[c]{@{}l@{}}\textit{If the participant's response to the question on the amount of requirements changes he/she received as a percentage of}\\\textit{ total work items indicated more than 50\%:}\\ It looks like you receive a considerable amount of requirements changes. \\ Does using agile methods enable you to respond to these requirement changes? If yes, how? \\ If not, why not?\\ \\ I\textit{f the participant's respond to the question on the amount of requirements changes he/she received} \\ \textit{as a percentage of total work items indicated less than 50\%:}\\ It looks like you don't receive a lot of requirements changes. So, why do you think practicing \\ agile is appropriate for your project?\end{tabular} & Open-ended                                                                                        \\ \bottomrule
\end{tabular}
\end{table*}

\subsection{Data Collection}
After the survey questionnaire was finalized, we sent the survey to 2 Research Fellows and 2 Ph.D. students in Australia and New Zealand who had software development industrial experience. This pilot run helped us to get feedback in terms of time for completion and any other aspects such as wording. We then distributed the survey via:

\begin{itemize}
    \item posting the survey link on professional software development groups and in our profiles in social media such as \emph{LinkedIn, Twitter,} and \emph{Facebook};
    \item sending the survey link to our known contacts in the software development industry; and
    \item \emph{Agile Alliance} posting the survey link on their \emph{LinkedIn, Twitter,} and \emph{Facebook} channels.
\end{itemize}

The survey was available online for a period of a month. 106 participants had started the survey and 42 participants fully completed the survey. Two responses from the completed responses were removed due to their feedback given at the end of the questionnaire mentioned that they completed the survey only to see what the questions were and that their answers were arbitrary.
We targeted only agile practitioners. We confirmed this by the participants' demographics and their self confirmation on practicing agile prior to begining the survey. As given in Table \ref{tab:participant_demograhics}, participants of our survey included developers, agile coaches/scrum masters, testers, business analysts, product owners, tech leads, and managers.

\begin{table*}[t]
\centering
\caption{Demographics of the Survey Participants (P\#: Participant ID; XT: Total Experience in Software Development Industry in Years; XTA: Total Experience in Agile in Years; XA: Experience with Agile Software Development Methods; SL: Sri Lanka; Au: Australia; USA: United States of America; CO: Colombia; NZ: New Zealand; SG: Singapore; UK: United Kingdom; NL: Netherlands; In: India)}
\label{tab:participant_demograhics}
\resizebox{\textwidth}{!}{%
\begin{tabular}{@{}rlllrrll@{}}
\toprule
\multicolumn{1}{l}{\textbf{P\#}} & \textbf{Age Group} & \textbf{Gender} & \textbf{Country} & \multicolumn{1}{l}{\textbf{XT}} & \multicolumn{1}{l}{\textbf{XTA}} & \textbf{XA}                                                                                                                                                                                                                              & \textbf{Role in the Project}                                                                             \\ \midrule
P1                                & 26 - 30            & Male            & SL               & 2.5                             & 2                                & Scrum, Kanban, ScrumBan                                                                                                                                                                                                                                                                  & Developer                                                                                                \\
P2                                & 20 - 25            & Male            & SL               & 2.5                             & 2.5                              & Scrum                                                                                                                                                                                                                                                                                    & Developer                                                                                                \\
P3                                & 26 - 30            & Female          & SL               & 2                               & 2                                & Scrum                                                                                                                                                                                                                                                                                    & Agile Coach/Scrum Master, Developer                                                                      \\
P4                                & 26 - 30            & Male            & SL               & 3.5                             & 3.5                              & Scrum, Kanban                                                                                                                                                                                                                                                                            & Developer                                                                                                \\
P5                                & 26 - 30            & Female          & SL               & 2                               & 1                                & Scrum                                                                                                                                                                                                                                                                                    & Tester                                                                                                   \\
P6                                & 26 - 30            & Female          & SL               & 2                               & 2                                & Scrum                                                                                                                                                                                                                                                                                    & Tester                                                                                                   \\
P7                                & 26 - 30            & Male            & SL               & 3                               & 1                                & Scrum                                                                                                                                                                                                                                                                                    & Developer                                                                                                \\
P8                                & 20 - 25            & Female          & AU               & 2                               & 1                                & Scrum, XP                                                                                                                                                                                                                                                                                & Developer                                                                                                \\
P9                                & 20 - 25            & Female          & SL               & 2.5                             & 2.5                              & Scrum, Feature Driven Development                                                                                                                                                                                                                                                        & Tester                                                                                                   \\
P10                               & 20 - 25            & Male            & SL               & 5                               & 8                                & \begin{tabular}[c]{@{}l@{}}Scrum, XP, Scrum XP combo, Kanban, Crystal, \\ Feature Driven Development, \\ Dynamic System Development\end{tabular}                                                                                                                                            & Agile Coach/Scrum Master                                                                                 \\
P11                               & 26 - 30            & Female          & SL               & 3.5                             & 2.5                              & Scrum                                                                                                                                                                                                                                                                                    & Tester                                                                                                   \\
P12                               & 26 - 30            & Female          & SL               & 2.5                             & 2.5                              & Scrum                                                                                                                                                                                                                                                                                    & Business Analyst                                                                                         \\
P13                               & 41- 45             & Male            & USA              & 20                              & 10                               & Not specified                                                                                                                                                                                                                                                                        & Developer                                                                                                \\
P14                               & 31 - 35            & Female          & CO               & 9                               & 3.5                              & Scrum                                                                                                                                                                                                                                                                                    & Business Analyst                                                                                         \\
P15                               & 31 - 35            & Male            & SL               & 6                               & 1                                & Scrum, Kanban                                                                                                                                                                                                                                                                            & Developer                                                                                                \\
P16                               & 31 - 35            & Male            & SL               & 10                              & 10                               & Scrum                                                                                                                                                                                                                                                                                    & \begin{tabular}[c]{@{}l@{}}Agile Coach/Scrum Master, \\ Product Owner, Developer, Tech Lead\end{tabular} \\
P17                               & 26 - 30            & Male            & AU               & 4                               & 2                                & Scrum                                                                                                                                                                                                                                                                                    & Developer                                                                                                \\
P18                               & 26 - 30            & Female          & SL               & 2.5                             & 2.5                              & Scrum                                                                                                                                                                                                                                                                                    & Agile Coach/Scrum Master                                                                                 \\
P19                               & 31 - 35            & Male            & NZ               & 10                              & 7                                & Scrum, XP, Kanban                                                                                                                                                                                                                                                                        & Agile Coach/Scrum Master                                                                                 \\
P20                               & 46 - 50            & Male            & AU               & 16                              & 5                                & Scrum                                                                                                                                                                                                                                                                                    & Product Owner                                                                                            \\
P21                               & 26 - 30            & Female          & SL               & 2                               & 2                                & Scrum                                                                                                                                                                                                                                                                                    & Tester                                                                                                   \\
P22                               & 26 - 30            & Female          & AU               & 3                               & 2                                & Scrum, Kanban                                                                                                                                                                                                                                                                            & Developer                                                                                                \\
P23                               & 36 - 40            & Male            & SL               & 12                              & 8                                & Kanban                                                                                                                                                                                                                                                                                   & Developer                                                                                                \\
P24                               & 26 - 30            & Male            & SL               & 5                               & 2.5                              & Scrum, Kanban, ScrumBan                                                                                                                                                                                                                                                                  & Tester                                                                                                   \\
P25                               & Above 50           & Male            & AU               & 30                              & 15                               & \begin{tabular}[c]{@{}l@{}}Feature Driven Development, \\ Dynamic System Development, \\ Primarily the Values \& Principles of \\ The Agile Manifesto \\ \end{tabular} & Manager                                                                                                  \\
P26                               & 26 - 30            & Female          & NZ               & 1                               & 1                                & Scrum                                                                                                                                                                                                                                                                                    & Tester                                                                                                   \\
P27                               & 31 - 35            & Male            & SL               & 8                               & 1                                & Scrum, Kanban                                                                                                                                                                                                                                                                            & Developer                                                                                                \\
P28                               & 20 - 25            & Male            & AU               & 30                              & 20                               & Scrum, XP, Scrum XP combo, Kanban                                                                                                                                                                                                                                                        & Product Owner, Manager                                                                                   \\
P29                               & 26 - 30            & Female          & SG               & 2                               & 2                                & Scrum                                                                                                                                                                                                                                                                                    & Developer                                                                                                \\
P30                               & 26 - 30            & Male            & SG               & 2                               & 1                                & Crystal                                                                                                                                                                                                                                                                                  & Developer                                                                                                \\
P31                               & 31 - 35            & Male            & UK               & 3                               & 3                                & Scrum, Kanban, SAFe                                                                                                                                                                                                                                                                      & Business Analyst                                                                                         \\
P32                               & 41- 45             & Male            & USA              & 23                              & 15                               & Scrum, XP, Kanban                                                                                                                                                                                                                                                                        & Developer, Manager                                                                                       \\
P33                               & 36 - 40            & Male            & NL               & 10                              & 5                                & Scrum                                                                                                                                                                                                                                                                                    & Product Owner                                                                                            \\
P34                               & 31 - 35            & Female          & SL               & 9                               & 5                                & Scrum, Kanban                                                                                                                                                                                                                                                                            & Agile Coach/Scrum Master                                                                                 \\
P35                               & 20 - 25            & Female          & IN               & 3.5                             & 3                                & Scrum, XP, Kanban, Feature Driven Development                                                                                                                                                                                                                                            & Developer                                                                                                \\
P36                               & 31 - 35            & Male            & SL               & 9                               & 4                                & Scrum                                                                                                                                                                                                                                                                                    & Tester                                                                                                   \\
P37                               & 41- 45             & Female          & NZ               & 25                              & 5                                & Scrum, Kanban                                                                                                                                                                                                                                                                            & Product Owner                                                                                            \\
P38                               & 31 - 35            & Male            & SL               & 8.5                             & 4                                & Scrum, Kanban, Feature Driven Development                                                                                                                                                                                                                                                & Agile Coach/Scrum Master                                                                                 \\
P39                               & 31 - 35            & Female          & IN               & 13                              & 2                                & Scrum                                                                                                                                                                                                                                                                                    & \begin{tabular}[c]{@{}l@{}}Agile Coach/Scrum Master, \\ Product Owner\end{tabular}                       \\
P40                               & 26 - 30            & Male            & SL               & 4                               & 4                                & Scrum                                                                                                                                                                                                                                                                                    & Developer                                                                                                \\ \bottomrule
\end{tabular}%
}
\end{table*}

Participants had 1 to 30 years of total experience in the software development industry and their agile experience ranged from 1 - 20 years in Scrum, Kanban, ScrumBan, XP, Scrum XP combo, Crystal, Feature Driven Development, Dynamic System Development, and SAFe. One of the participants (P13) had not mentioned the agile methods that he had experience in and another participant (P28) had selected his age group as 20-25 even though his total experience in software development industry was indicated as 30 years. As we did not collect any contact information of the participants, we were not able to follow up to find the true information in both cases.

The majority of the participants (N=26) were from Asia (N(Sri Lanka)=22, N(Singapore)=2, N(India)=2). 9 participants were from Oceania (N(Australia)=6, N(New Zealand)=3), 3 participants were from North America (N(United States of America)=2, N(Colombia)=1), and 2 participants were from Europe (N(United Kingdom)=1, N(Netherlands)=1).

\subsection{Data Analysis}
Quantitative data were analysed using \emph{Qualtrics} and \emph{Microsoft Excel} while qualitative data were analysed using \emph{Microsoft Excel} and \emph{MAXQDA}. Qualitative data followed the coding approach given in GT \cite{Strauss1998BasicsTechniques.} where concepts and categories were generated through constant comparison across different dimensions such as separate/hybrid and degree of proper use.



\begin{figure}[b]
    \centering
    \includegraphics[width=\columnwidth]{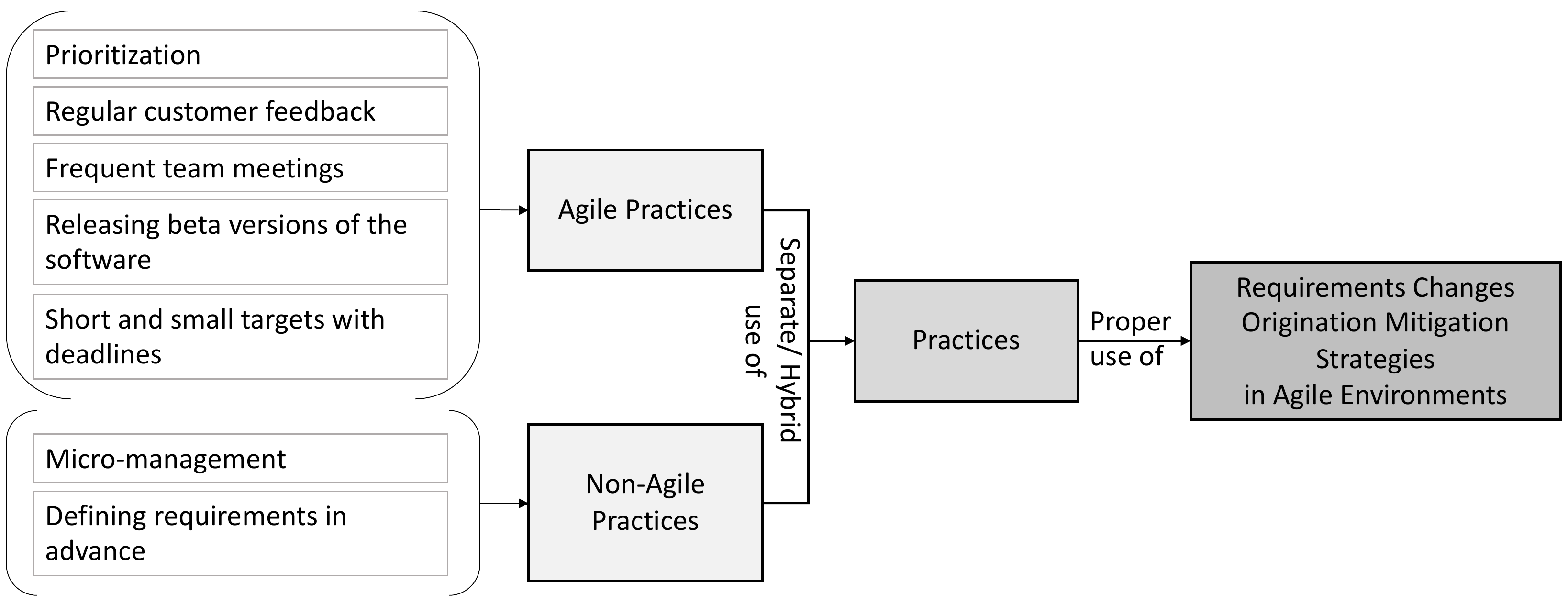}
    \caption{Emergence of Category ``Requirements Changes Origination Mitigation Strategies'' using Theory Data Analysis}
    \label{fig:GT}
\end{figure}

An example of the GT analysis is shown in Fig. \ref{fig:GT}. The \textbf{codes: prioritization, regular customer feedback, frequent team meetings, releasing beta versions of the software,} and \textbf{short and small targets with deadlines} led to the \textbf{concept: agile practices}, and the \textbf{codes: micro-management, and defining requirements in advance} led to the \textbf{concept non-agile practices}. These concepts led to the \textbf{sub category: practices} through its \textbf{dimension of separate/hybrid use (axial coding)}. Finally the \textbf{category: requirements changes origination strategies} was emerged from the sub category.

%% file: sections/type.tex
The types of RCs we found through interviews and literature: \emph{Functional requirement (FR) addition, FR modification, FR deletion, FR bug fix, FR combination, Non-functional requirement (NFR) addition, NFR modification, NFR deletion, NFR bug fix, NFR combination}, and \emph{FR-NFR combination} were given as options for participants to choose as the types of RCs they receive. 

In addition, an \emph{other} option was also given in the survey for the participants to provide if any other types exist apart from the types we have given them to choose. Participants were able to choose multiple types. The rest of the survey questions depended on the choices in this question as they only had to answer the other questions regarding RC types that they selected in this question.

We provided the below acronyms, definitions, and examples for the participants to use as a guide to answer this question. Additionally, we assumed that \emph{deletion} and \emph{combination} as understandable terms.

\begin{itemize}
    \item \textbf{FR:} Functional Requirement
    \item \textbf{NFR:} Non-Functional Requirement
    \item \textbf{Bug Fix:} Correction in the codebase
    \item \textbf{Addition:} A new requirement arising due to a change in an existing requirement
    \item \textbf{Modification:} Modifying an actual requirement. E.g., Modify the actual user story/split the user story/ change user story partially
\end{itemize}

\begin{figure}[t]
        \centering
        \includegraphics[, width=\columnwidth]{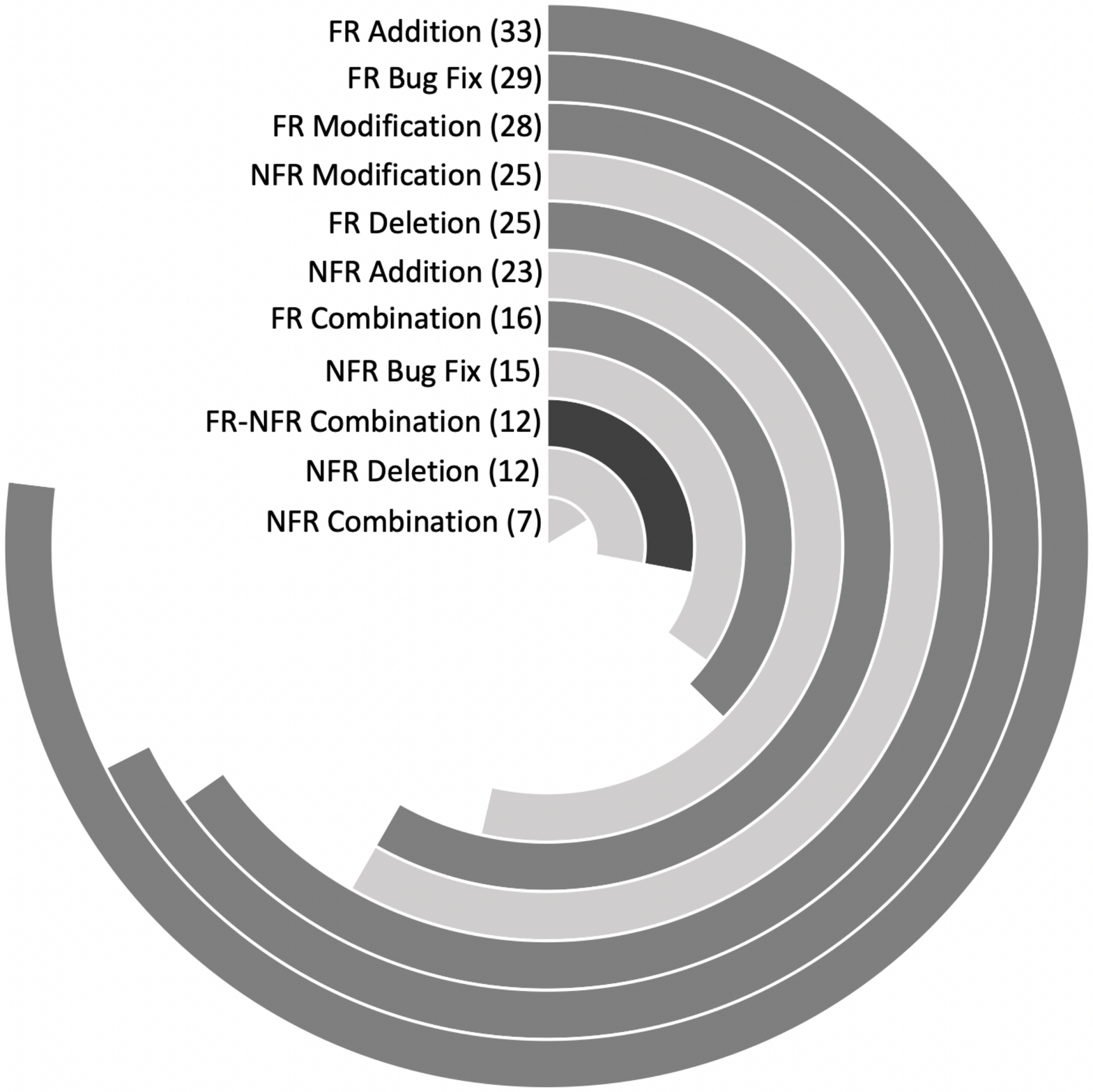}
        \caption{Types of Requirements Changes (Number of Responses}
        \label{fig:types}
    \end{figure}

Fig. \ref{fig:types} shows the different types of RCs reported. The top most received RC type as reported by the participants was \emph{FR addition} (N=33). It was followed by \emph{FR bug fix} (N=29), and \emph{FR modification} (N=28). Same amount of participants (N=25) reported that they receive \emph{FR deletion} and \emph{NFR modification}. 23 out the 40 participants reported that they receive \emph{NFR additions} as RCs. \emph{FR combination} (N=16), \emph{NFR bug fix} (N=15), \emph{NFR deletion} (N=12), \emph{FR-NFR combination} (N=12), and \emph{NFR combination} (N=7) were also received by the participants as RCs. In addition, participant P12 (Business Analyst) reported that she had experienced a type of an RC called ``transition requirements addition'', which we considered as either a functional or a non-functional RC.

FRs change was the most commonly received RC type in agile environments in terms of additions, bug fixes, modifications, and deletions (N(Total FRs)=131). Also, a substantial amount of RCs as NFR modifications and additions can be seen in the results (N(NFRs)=82). 

%% file: sections/reason.tex
We define root causes of RC origination as \emph{reasons}. \emph{Functional enhancement, design improvement, missing requirements, requirements clarification, bug (Error arising in the codebase), scope reduction, erroneous requirements, resolving conflicts, product strategy, need for refactoring, outstanding technical debt, obsolete functionality, redundant functionality}, and \emph{other} were given as the reasons for RC origination in the form of multiple choices in the survey for the participants to select from. These reasons were derived from the our interview-based study data and from existing literature. We categorized these reasons as \emph{software-centric} and \emph{human-centric} reasons as given in the Table \ref{tab:reason types}. By \emph{software-centric} we mean the reasons that are directly linked to the software being developed, and by \emph{human-centric} we mean the reasons that are caused due to the approaches taken by the humans involved in the project.

\begin{table}[b]
\caption{Categories of Reasons for Requirements Changes Origination}
\label{tab:reason types}
\resizebox{\columnwidth}{!}{%
\begin{tabular}{@{}lll@{}}
\toprule
\multicolumn{1}{c}{\textbf{Category}}      & \multicolumn{1}{c}{\textbf{Reason}}          & \multicolumn{1}{c}{\textbf{Responses}} \\ \midrule
\multirow{6}{*}{Software-centric} & Functional enhancement              & 32                                      \\
                                  & Design improvement                  & 28                                      \\
                                  & Bug (Error arising in the codebase) & 21                                      \\
                                  & Errorneous requirements             & 17                                      \\
                                  & Redundant functionality             & 9                                       \\
                                  & Obsolete functionality              & 9                                       \\ \midrule
\multirow{7}{*}{Human-centric}    & Missing requirements                & 27                                      \\
                                  & Requirements clarification          & 25                                      \\
                                  & Scope reduction                     & 19                                      \\
                                  & Resolving conflicts                 & 16                                      \\
                                  & Product strategy                    & 15                                      \\
                                  & Need for refactoring                & 11                                      \\
                                  & Outstanding technical debt          & 10                                      \\ \bottomrule
\end{tabular}%
}
\end{table}

As shown in Table \ref{tab:reason types}, most participants (N=32) reported that the reason for RC origination is functional enhancement. Design improvement (N=28), missing requirements (N=27), requirements clarification (N=25), and bug (N=21) were chosen as the reasons for RCs to occur by more than half of the participants. Scope reduction (N=19), erroneous requirements (N=17), resolving conflicts (N=16), product strategy (N=15), need for refactoring (N=11), and outstanding technical debt (N=10) were selected by more than or equal to 25\% of the participants. Obsolete functionality and redundant functionality were selected at the same count (N=9) by the participants which are the least common reasons as reported.

Considering the responses which are above half of the sample size ($>$N=20), 3 out of 5 reasons (functional enhancement, design improvement, and bug) for RC occurrence are software-centric. The other 2 reasons (missing requirements, and requirements clarification,) are human-centric.

Taking the rest of the responses ($<$N=20) into consideration, 5 out of 8 reasons (scope reduction, resolving conflicts, product strategy, need for refactoring, and outstanding technical debt) are human-centric. The other 3 reasons (erroneous requirements, obsolete functionality, and redundant functionality) are software-centric.


Apart from the above mentioned reasons, we found the following additional reasons reported for the origination of RCs in open-ended responses:

\begin{itemize}
    \item Inadequate communication
    \item Inadequate documentation
    \item Rushed analysis when defining requirements
    \item Wrong set of initial requirements
\end{itemize}

\textbf{Inadequate communication:}
\textit{``Individuals and interactions over processes and tools''} is a core value in agile \cite{Beck2001ManifestoDevelopment}. In order to have better interactions, better communication is expected to prevail. However, if communication among the stakeholders is at an inadequate level, origination of RCs can be expected as reported by the participants. In this case, it is required for the team to be mindful and overcome this issue so that unexpected RCs do not originate. 

\textbf{Inadequate documentation:}
\textit{Even though agile encourages ``working software over comprehensive documentation''} \cite{Beck2001ManifestoDevelopment}, as reported by the participants, inadequate documentation is a reason for some RC origination. Documenting \emph{the requirements} to a sufficient level is therefore recommended.

\textbf{Rushed analysis when defining requirements:}
Even though the root causes are unknown for rushing the analysis of requirements, it causes the RCs to originate later:
\begin{center}
    \textit{``The people defining what the software should do have either rushed their analysis...`` - P25 [Manager]}
\end{center}

\textbf{Wrong set of initial requirements:}
Having the wrong set of initial requirements unsurprisingly also causes RCs. RCs are necessary to redirect the software development in the correct direction so as to meet the actual customer requirements:
\begin{center}
    \textit{``If the requirements change, that means the previous set of requirements was wrong, and who wants to build the wrong thing?'' - P32 [Developer/Manager]''}
\end{center}

%% file: sections/source.tex
We define human and non-human artefacts which lead to origination of RCs as \emph{sources}. We found that RCs originate from \emph{defect reports, individual developer's detailed analysis, marketing team, product backlog review, technical team discussion, user reviews}, and from \emph{user-support discussions}. We categorized these sources as \emph{in-team software-centric, in-team human-centric, out-team software-centric,} and \emph{out-team human-centric} as given in Table \ref{tab:source_cat}. 
By \emph{in-team} we mean the agile team, by \emph{out-team} we mean stakeholders outside the agile team, by \emph{software-centric} we mean the non-human artefacts directly linked to the software being developed, and by \emph{human-centric} we mean the approaches taken in terms of activities by the humans involved in the software development process. 

Our findings are shown in Table \ref{tab:source}. We discuss the most provided responses for each type of RC in the sub-sections below. We used a likert scale of less than average, average, and more than average as the scale to answer this question.

\begin{table}[]
\centering
\caption{Categories of Sources of Requirements Changes}
\label{tab:source_cat}
\begin{tabular}{@{}ll@{}}
\toprule
\multicolumn{1}{c}{\textbf{Source}}                & \multicolumn{1}{c}{\textbf{Category}} \\ \midrule
Defect reports                           & In-team software-centric    \\
Individual developer's detailed analysis & In-team software-centric    \\
Product backlog reviews                   & In-team human-centric       \\
Technical team discussions                & In-team human-centric       \\ \midrule
User reviews                             & Out-team software-centric   \\
Marketing team                           & Out-team human-centric      \\
User-support discussions                 & Out-team human-centric      \\ \bottomrule
\end{tabular}
\end{table}

\subsubsection{Sources of Functional Requirements Changes}

\begin{table*}[t]
\fontsize{14}{16}\selectfont
\caption{Sources of Requirements Changes (\includegraphics[scale=0.2]{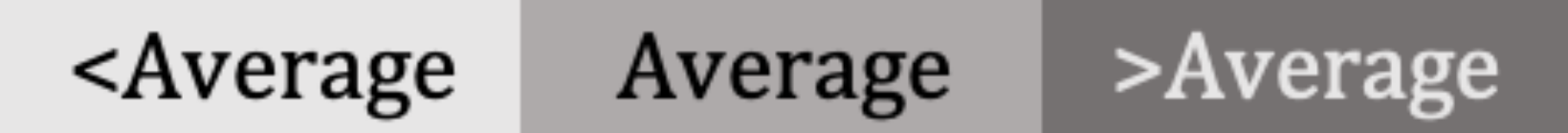}; FR: Functional Requirement; NFR: Non-Functional Requirement)}
\label{tab:source}
\resizebox{\textwidth}{!}{%
\begin{tabular}{@{}llllllllllllllllllllll@{}}
\toprule
\multicolumn{1}{c}{\textbf{}}                & \multicolumn{3}{c}{\textbf{Defect Reports}}                                                             & \multicolumn{3}{c}{\textbf{\begin{tabular}[c]{@{}c@{}}Individual Developer's \\ Detailed Analysis\end{tabular}}} & \multicolumn{3}{c}{\textbf{\begin{tabular}[c]{@{}c@{}}Product Backlog\\ Reviews\end{tabular}}}          & \multicolumn{3}{c}{\textbf{\begin{tabular}[c]{@{}c@{}}Technical Team \\Discussions\end{tabular}}}                                                 & \multicolumn{3}{c}{\textbf{User Reviews}}                                                               & \multicolumn{3}{c}{\textbf{Marketing Team}} & \multicolumn{3}{c}{\textbf{\begin{tabular}[c]{@{}c@{}}User-Support \\Discussions\end{tabular}}} \\ \midrule
\multicolumn{22}{l}{\textbf{Functional Requirements Changes}}    \\ \midrule
FR Addition                                  & \multicolumn{3}{l}{\includegraphics[width=5.2cm,height=0.6cm]{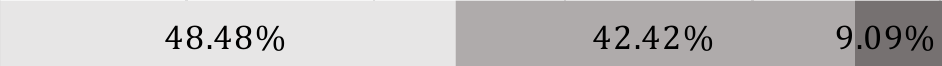}}                                                                           & \multicolumn{3}{l}{\includegraphics[width=5.2cm,height=0.6cm]{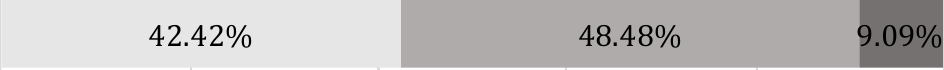}}                                                                                    & \multicolumn{3}{l}{\includegraphics[width=5.2cm,height=0.6cm]{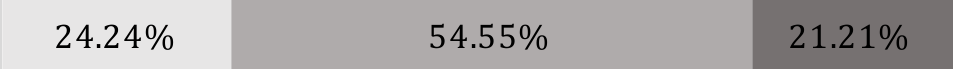}}                                                                          & \multicolumn{3}{l}{\includegraphics[width=5.2cm,height=0.6cm]{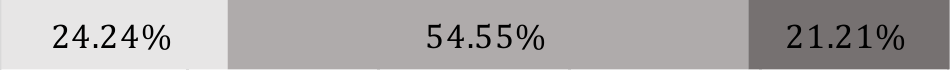}}                                                  & \multicolumn{3}{l}{\includegraphics[width=5.2cm,height=0.6cm]{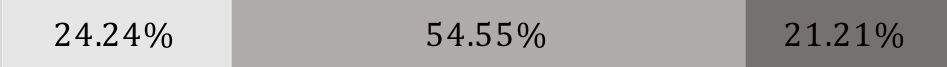}}                                                            & \multicolumn{3}{l}{\includegraphics[width=5.2cm,height=0.6cm]{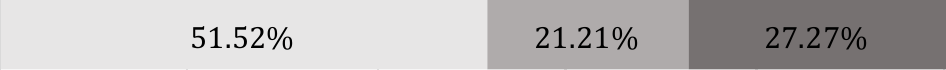}}                                                                                    & \multicolumn{3}{l}{\includegraphics[width=5.2cm,height=0.6cm]{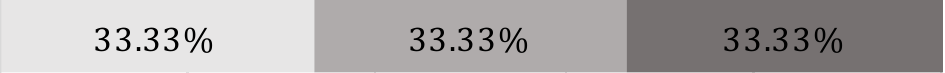}}   \\
FR Modification                              & \multicolumn{3}{l}{\includegraphics[width=5.2cm,height=0.6cm]{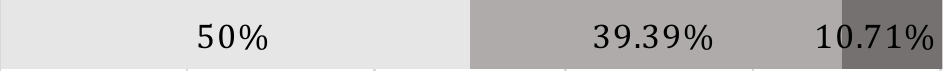}}                                                                           & \multicolumn{3}{l}{\includegraphics[width=5.2cm,height=0.6cm]{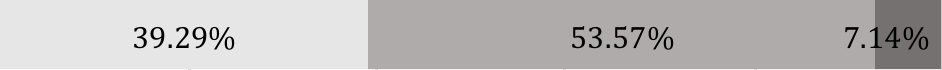}}                                                                                     & \multicolumn{3}{l}{\includegraphics[width=5.2cm,height=0.6cm]{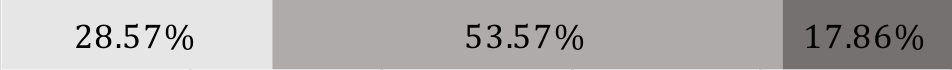}}                                                                         & \multicolumn{3}{l}{\includegraphics[width=5.2cm,height=0.6cm]{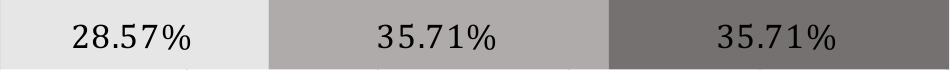}}                                                  & \multicolumn{3}{l}{\includegraphics[width=5.2cm,height=0.6cm]{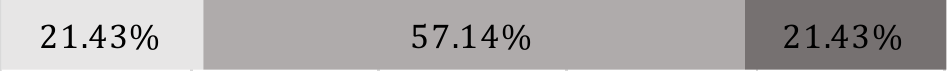}}                                                             & \multicolumn{3}{l}{\includegraphics[width=5.2cm,height=0.6cm]{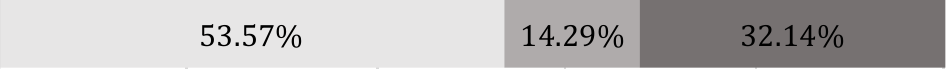}}                                                                                    & \multicolumn{3}{l}{\includegraphics[width=5.2cm,height=0.6cm]{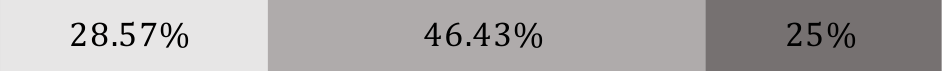}}    \\
FR Deletion                                  & \multicolumn{3}{l}{\includegraphics[width=5.2cm,height=0.6cm]{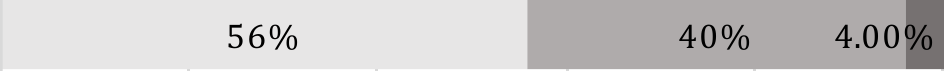}}                                                                           & \multicolumn{3}{l}{\includegraphics[width=5.2cm,height=0.6cm]{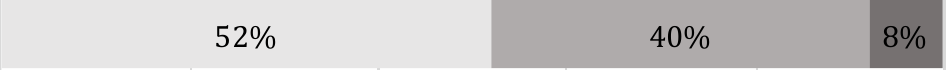}}                                                                                     & \multicolumn{3}{l}{\includegraphics[width=5.2cm,height=0.6cm]{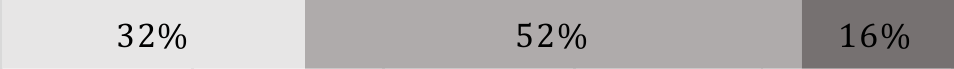}}                                                                   & \multicolumn{3}{l}{\includegraphics[width=5.2cm,height=0.6cm]{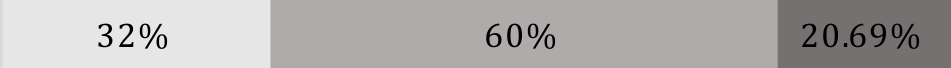}}                                                  & \multicolumn{3}{l}{\includegraphics[width=5.2cm,height=0.6cm]{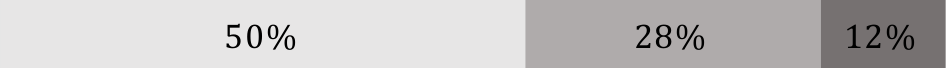}}                                                                   & \multicolumn{3}{l}{\includegraphics[width=5.2cm,height=0.6cm]{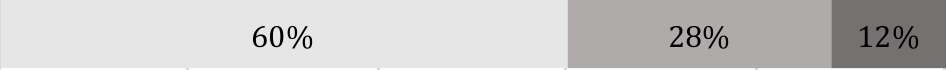}}                                                                                    & \multicolumn{3}{l}{\includegraphics[width=5.2cm,height=0.6cm]{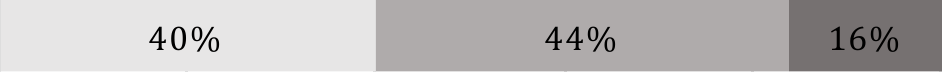}}   \\
FR Bug Fix                                   & \multicolumn{3}{l}{\includegraphics[width=5.2cm,height=0.6cm]{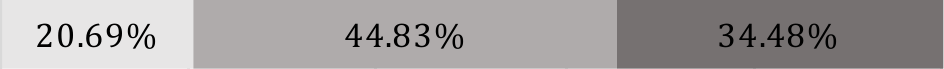}}                                                                           & \multicolumn{3}{l}{\includegraphics[width=5.2cm,height=0.6cm]{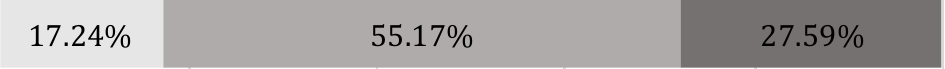}}                                                                            & \multicolumn{3}{l}{\includegraphics[width=5.2cm,height=0.6cm]{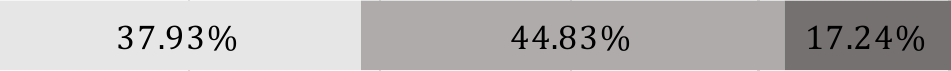}}                                                  & \multicolumn{3}{l}{\includegraphics[width=5.2cm,height=0.6cm]{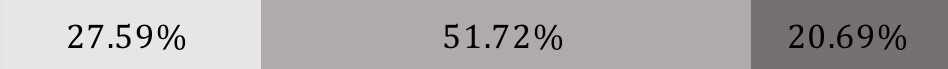}}                                                           & \multicolumn{3}{l}{\includegraphics[width=5.2cm,height=0.6cm]{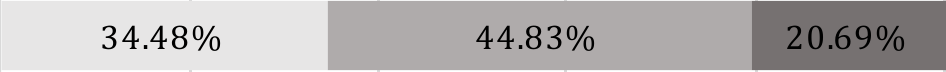}}                                                                                    & \multicolumn{3}{l}{\includegraphics[width=5.2cm,height=0.6cm]{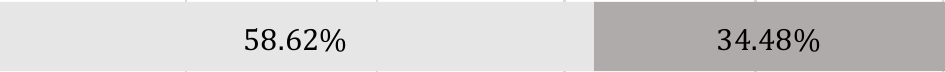}}                                                                           & \multicolumn{3}{l}{\includegraphics[width=5.2cm,height=0.6cm]{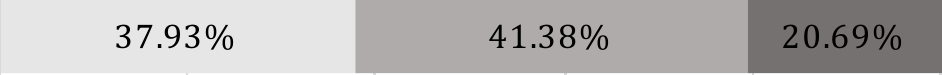}}    \\
FR Combination                               & \multicolumn{3}{l}{\includegraphics[width=5.2cm,height=0.6cm]{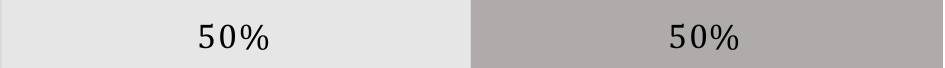}}                                                                           & \multicolumn{3}{l}{\includegraphics[width=5.2cm,height=0.6cm]{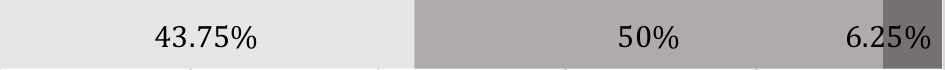}}                                                                                     & \multicolumn{3}{l}{\includegraphics[width=5.2cm,height=0.6cm]{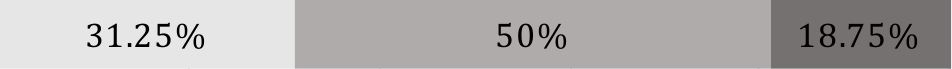}}                                                                      & \multicolumn{3}{l}{\includegraphics[width=5.2cm,height=0.6cm]{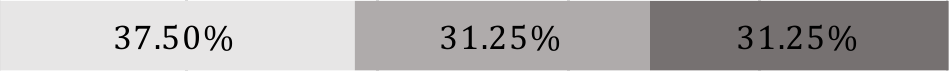}}                                                  & \multicolumn{3}{l}{\includegraphics[width=5.2cm,height=0.6cm]{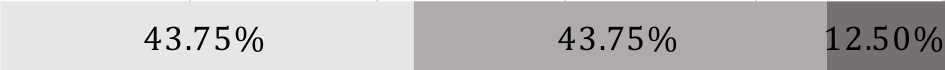}}                                                                & \multicolumn{3}{l}{\includegraphics[width=5.2cm,height=0.6cm]{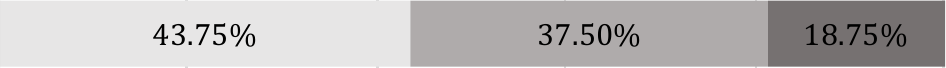}}                                                                                    & \multicolumn{3}{l}{\includegraphics[width=5.2cm,height=0.6cm]{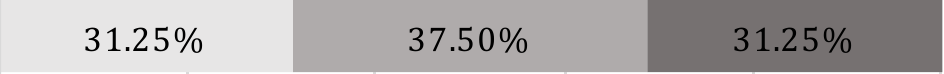}}  \\ \midrule
\multicolumn{22}{l}{\textbf{Non-Functional Requirements Changes}}    \\ \midrule
NFR Addition                                 & \multicolumn{3}{l}{\includegraphics[width=5.2cm,height=0.6cm]{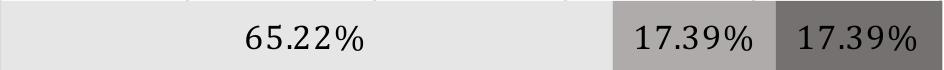}}                                                                                    & \multicolumn{3}{l}{\includegraphics[width=5.2cm,height=0.6cm]{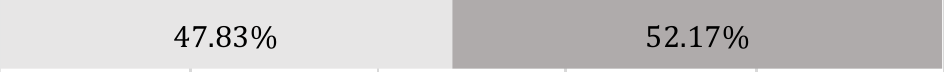}}                                                                                     & \multicolumn{3}{l}{\includegraphics[width=5.2cm,height=0.6cm]{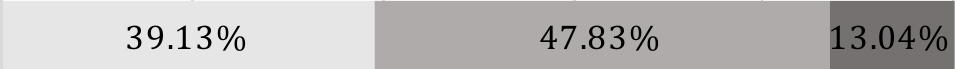}}                                                  & \multicolumn{3}{l}{\includegraphics[width=5.2cm,height=0.6cm]{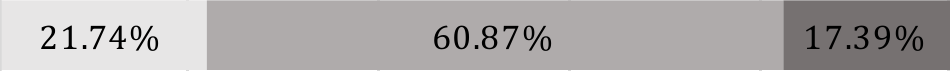}}                                                           & \multicolumn{3}{l}{\includegraphics[width=5.2cm,height=0.6cm]{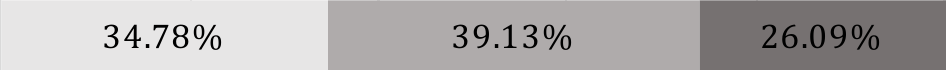}}                                                                           & \multicolumn{3}{l}{\includegraphics[width=5.2cm,height=0.6cm]{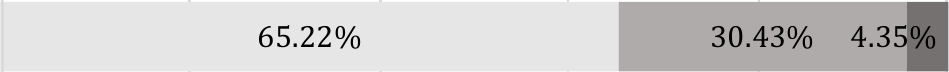}}                                                                           & \multicolumn{3}{l}{\includegraphics[width=5.2cm,height=0.6cm]{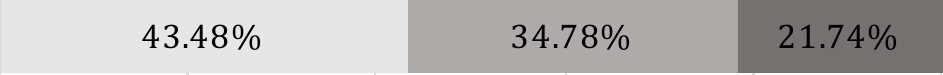}}   \\
NFR Modification                             & \multicolumn{3}{l}{\includegraphics[width=5.2cm,height=0.6cm]{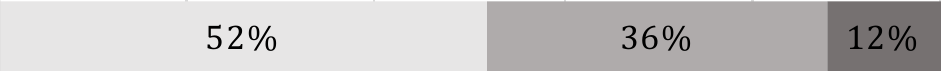}}                                                                                    & \multicolumn{3}{l}{\includegraphics[width=5.2cm,height=0.6cm]{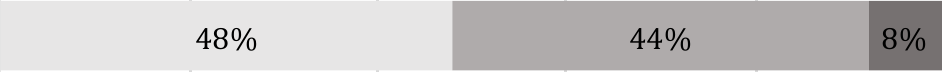}}                                                                                     & \multicolumn{3}{l}{\includegraphics[width=5.2cm,height=0.6cm]{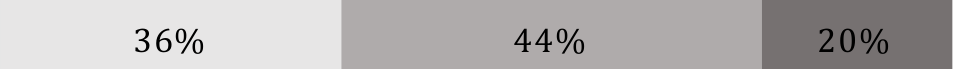}}                                                  & \multicolumn{3}{l}{\includegraphics[width=5.2cm,height=0.6cm]{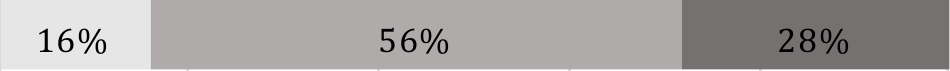}}                                                           & \multicolumn{3}{l}{\includegraphics[width=5.2cm,height=0.6cm]{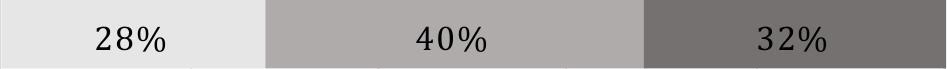}}                                                                           & \multicolumn{3}{l}{\includegraphics[width=5.2cm,height=0.6cm]{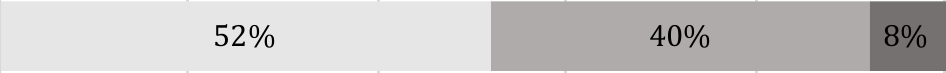}}                                                                           & \multicolumn{3}{l}{\includegraphics[width=5.2cm,height=0.6cm]{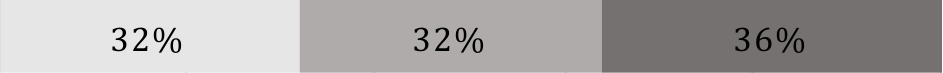}}    \\
NFR Deletion                                 & \multicolumn{3}{l}{\includegraphics[width=5.2cm,height=0.6cm]{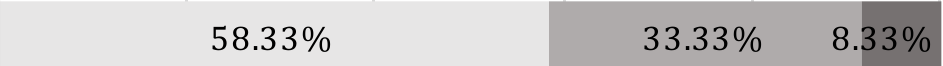}}                                                                                    & \multicolumn{3}{l}{\includegraphics[width=5.2cm,height=0.6cm]{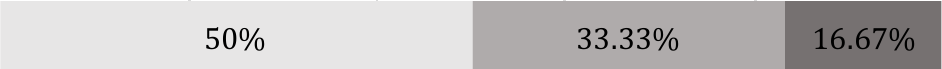}}                                                                                    & \multicolumn{3}{l}{\includegraphics[width=5.2cm,height=0.6cm]{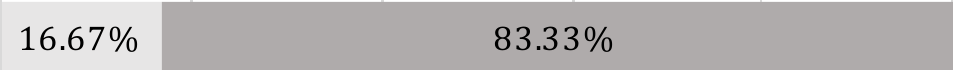}}                                                  & \multicolumn{3}{l}{\includegraphics[width=5.2cm,height=0.6cm]{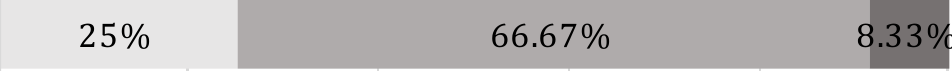}}                                                            & \multicolumn{3}{l}{\includegraphics[width=5.2cm,height=0.6cm]{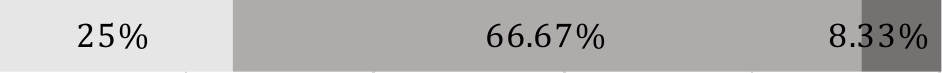}}                                                                           & \multicolumn{3}{l}{\includegraphics[width=5.2cm,height=0.6cm]{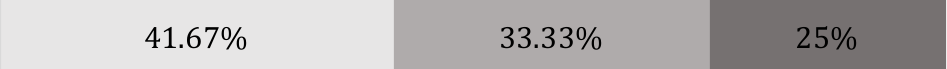}}                                                                           & \multicolumn{3}{l}{\includegraphics[width=5.2cm,height=0.6cm]{figures/source/support/SupportNFRD.pdf}}    \\
NFR Bug Fix                                  & \multicolumn{3}{l}{\includegraphics[width=5.2cm,height=0.6cm]{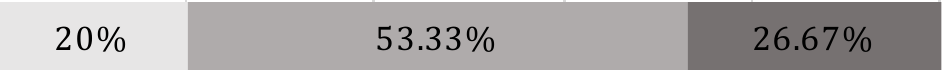}}                                                                                    & \multicolumn{3}{l}{\includegraphics[width=5.2cm,height=0.6cm]{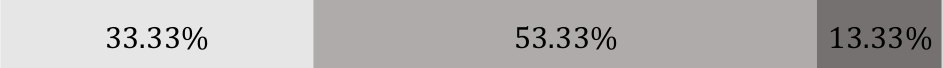}}                                                                                 & \multicolumn{3}{l}{\includegraphics[width=5.2cm,height=0.6cm]{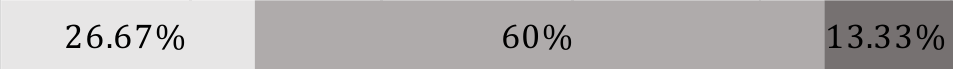}}                                                  & \multicolumn{3}{l}{\includegraphics[width=5.2cm,height=0.6cm]{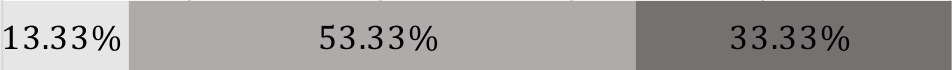}}                                                      & \multicolumn{3}{l}{\includegraphics[width=5.2cm,height=0.6cm]{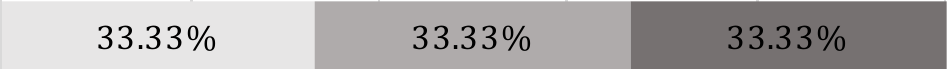}}                                                                           & \multicolumn{3}{l}{\includegraphics[width=5.2cm,height=0.6cm]{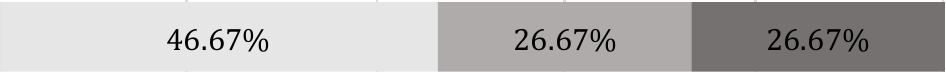}}                                                                           & \multicolumn{3}{l}{\includegraphics[width=5.2cm,height=0.6cm]{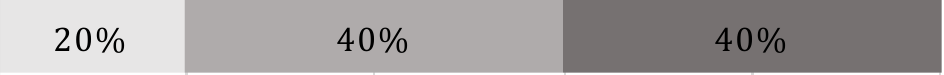}} \\
NFR Combination                              & \multicolumn{3}{l}{\includegraphics[width=5.2cm,height=0.6cm]{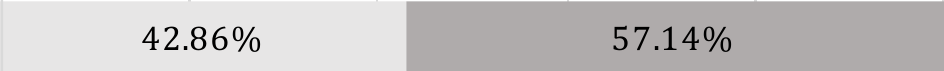}}                                                                                    & \multicolumn{3}{l}{\includegraphics[width=5.2cm,height=0.6cm]{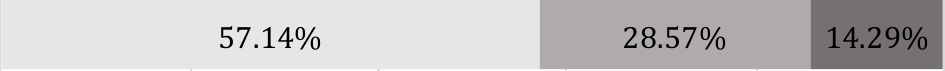}}                                                                   & \multicolumn{3}{l}{\includegraphics[width=5.2cm,height=0.6cm]{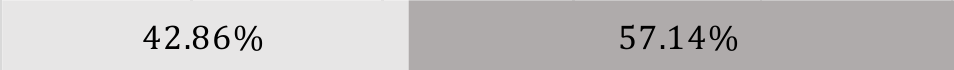}}                                                  & \multicolumn{3}{l}{\includegraphics[width=5.2cm,height=0.6cm]{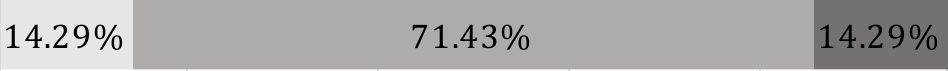}}                                                                    & \multicolumn{3}{l}{\includegraphics[width=5.2cm,height=0.6cm]{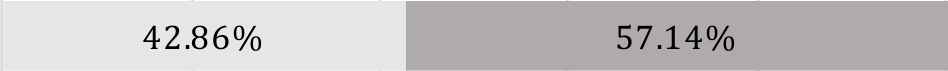}}                                                                           & \multicolumn{3}{l}{\includegraphics[width=5.2cm,height=0.6cm]{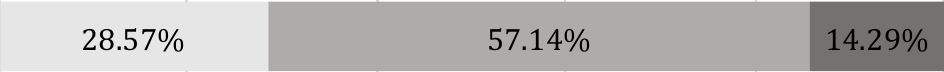}}                                                                           & \multicolumn{3}{l}{\includegraphics[width=5.2cm,height=0.6cm]{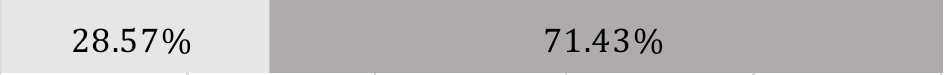}}                                                                                     \\ \midrule
FR-NFR Combination                           & \multicolumn{3}{l}{\includegraphics[width=5.2cm,height=0.6cm]{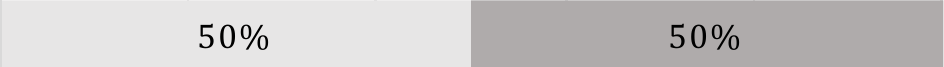}}                                                                           & \multicolumn{3}{l}{\includegraphics[width=5.2cm,height=0.6cm]{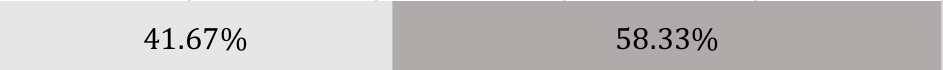}}                                                                    & \multicolumn{3}{l}{\includegraphics[width=5.2cm,height=0.6cm]{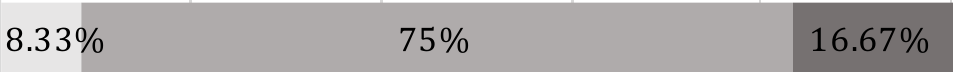}}                                                  & \multicolumn{3}{l}{\includegraphics[width=5.2cm,height=0.6cm]{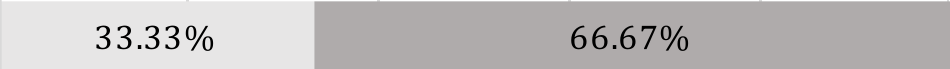}}                                                                   & \multicolumn{3}{l}{\includegraphics[width=5.2cm,height=0.6cm]{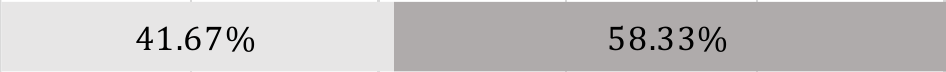}}                                                                           & \multicolumn{3}{l}{\includegraphics[width=5.2cm,height=0.6cm]{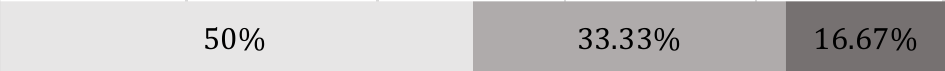}}                                                                           & \multicolumn{3}{l}{\includegraphics[width=5.2cm,height=0.6cm]{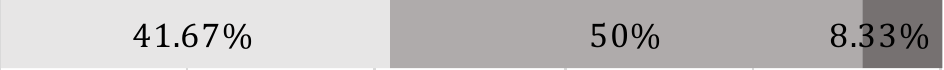}}                                                                                         \\ \bottomrule
\end{tabular}%
}
\end{table*}


\textbf{FR Addition:}
FR additions occur from \textit{defect reports} (48.48\%), and \textit{marketing team} (51.52\%), less than average of time as reported by the majority of participants. FR additions occurring from \textit{user-support discussions} was equally found as same number of responses (33.33\% each) were given for the options: ``less than average, average'', and ``more than average''. Due to having the same number of responses for both options, it is unable to suggest the exact frequency of FR additions which occur from \textit{user-support discussions}. FR additions occur from \textit{individual developer's detailed analysis} (48.48\%), \textit{product backlog reviews} (54.55\%), \textit{technical team discussions} (54.55\%), and \textit{user reviews} (54.55\%), on an average mostly. 

\textbf{FR Modification:}
It can be seen that FR modifications originate in \textit{technical team discussions} (35.71\% each) more than average and on average, in \textit{individual developer's detailed analysis} (53.57\%), \textit{product backlog reviews} (53.57\%), \textit{user reviews} (57.14\%), and \textit{user-support discussions} (46.43\%), it is on average, and in \textit{defect reports} (50\%) and \textit{marketing team} (53.57\%) it is less than average. 

\textbf{FR Deletion:}
Our findings indicate that, FR deletions originate less than average from \textit{defect reports} (56\%), \textit{individual developer's detailed analysis} (52\%), \textit{marketing team} (60\%), and in \textit{user reviews} (50\%). Additionally, they originate from \textit{product backlog reviews} (52\%), \textit{technical team discussions} (60\%), and \textit{user-support discussions} (44\%) on average.

\textbf{FR Bug Fix:}
FR bug fixes can be found less than average from \textit{marketing team} (58.62\%) and they can be found in \textit{defect reports} (44.83\%), \textit{individual developer's detailed analysis} (55.17\%), \textit{product backlog reviews} (44.83\%), \textit{technical team discussions} (51.72\%), \textit{user reviews} (44.83\%), and \textit{user-support discussions} (41.38\%) on average.

\textbf{FR Combination:}
Our findings show that FR combinations occur in \textit{individual developer's detailed analysis }(50\%), \textit{product backlog reviews} (50\%), and in \textit{user-support discussions} (37.5\%) on an average. Moreover, occurrence of FR combinations is less than average from \textit{marketing team} (43.75\%) and \textit{technical team discussions} (37.5\%). However, it is not clear whether FR combinations occur in \textit{defect reports} and \textit{user reviews} on an average or less than average as same amount of responses were found (defect reports=50\% each, user reviews=43.75\% each).

\subsubsection{Sources of Non-Functional Requirements Changes}

\textbf{NFR Addition:}
As reported by the majority of participants, NFR additions originate from \textit{individual developer's detailed analysis} (52.17\%), \textit{product backlog reviews} (47.83\%), \textit{technical team discussions} (60.87\%), and \textit{user reviews} (39.13\%) on an average and they originate less than average from \textit{marketing team} (65.22\%), and \textit{user-support discussions} (43.48\%). Same number of participants (17.39\% each) reported that NFR additions originate from defect reports on an average and more than average. However, as it is either ``average'' or ``more than average'', this suggests that NFR additions occur from \textit{defect reports} in a considerable amount.

\textbf{NFR Modification:}
NFR modifications originate from: \textit{user-support discussions} (36\%) more than average, \textit{product backlog reviews} (44\%), \textit{technical team discussions} (56\%), and \textit{user reviews} (40\%) on an average, and \textit{defect reports} (52\%), \textit{individual developer's detailed analysis} (48\%), and \textit{marketing team} (52\%) less than average. 

\textbf{NFR Deletion:}
NFR  deletions originate from \textit{product backlog reviews} (83.33\%), \textit{technical team discussions} (56\%), and \textit{user-support discussions} (66.67\%) on an average and they originate less than average from \textit{defect reports} (58.33\%), \textit{individual developer's detailed analysis} (50\%), and from \textit{marketing team} (41.67\%). However, it is required to look further the probability of NFR deletions to originate from \textit{user reviews} (out-team software-centric) as the same number of responses (33.33\% each) were found from our survey. 

\textbf{NFR Bug Fix:}
NFR bug fixes rise from \textit{defect reports} (53.33\%), \textit{individual developers' detailed analysis} (53.33\%), \textit{product backlog reviews} (60\%), \textit{technical team discussions} (53.33\%), and from \textit{user reviews} (53.33\%) on average and from less than average from \textit{marketing team} (46.67\%). However, it is unclear whether they originate on average or more than average from \textit{user-support discussions} as same number of responses were found (40\% each).

\textbf{NFR Combination:}
NFR combinations originate from \textit{defect reports} (57.14\%), \textit{marketing team} (57.14\%), \textit{product backlog reviews} (57.14\%), \textit{technical team discussions} (71.43\%), \textit{user reviews} (57.14\%), and from \textit{user-support discussions} (71.43\%). In addition, NFR combinations originate less than average from \textit{individual developer's detailed analysis} (57.14\%). 

\subsubsection{Sources of Combination of Functional and Non-Functional Requirements Changes:}
FR-NFR combinations originate from \textit{individual developer's detailed analysis} (58.33\%), \textit{product backlog reviews} (75\%), \textit{technical team discussions} (66.67\%), \textit{user reviews} (58.33\%), and \textit{user-support discussions} (50\%) on an average and from \textit{marketing team} (50\%) less than average. However, \textit{defect reports} is a gray source where half of the participants reported that it is on average and the other half reported that it is less than average for FR-NFR combinations to originate from defect reports. It is required to look into this further to know whether it is on average or less than average. 

\subsubsection{Sources of Requirements Changes According to the Source Category}
We summarize the above results according to the source category.

\textbf{FR Addition:} 3 out of the 4 sources (individual developer's detailed analysis, product backlog reviews, technical team discussions) from which FR additions occur on average, are in-team human-centric sources and the other (user reviews) is an out-team software-centric source.

\textbf{FR Modification:} 1 out of the 5 sources (user-support discussions) where FR modifications originate on average is out-team human-centric, one is out-team software-centric (user reviews), one is in-team software-centric (individual developer's detailed analysis), and the other two (product backlog reviews and technical team discussions) are in-team human-centric sources. Therefore, this indicates that sources of FR modifications are human-centric mostly.

\textbf{FR Deletion:} 2 out of 3 (product backlog reviews, technical-team discussions) are in-team human-centric sources and the other source (user-support discussions) is an out-team human-centric source. However, all these sources are human-centric.

\textbf{FR Bug Fix:} 2 out of these 6 sources (defect reports, individual developer's detailed analysis) are in-team software-centric and the remaining sources are in-team human-cenric (product backlog reviews, technical team discussions), out-team software-centric (user reviews), and out-team human-centric (user-support discussions). This indicates that majority of sources of FR bug fixes are human-centric. 

\textbf{FR Combination:} Taking the sources which had majority of responses for option ``average'', one is in-team software-centric (individual developer's detailed analysis) and the other (user-support discussions) is out-team human-centric. The unclear sources (defect reports and user reviews) are in-team software-centric and out-team software-centric. However, this suggests that the sources of FR combinations are mostly software-centric.

\textbf{NFR Addition:} Out of the 4 sources which had ``average'' as the highest response, 3 are in-team human-centric (individual developer's detailed analysis, product backlog reviews, technical team discussions) sources and the other (user reviews) is an out-team software-centric source. Similar to the other types of requirements changes, the majority of the sources for NFR additions are human-centric.
 
\textbf{NFR Modification:} The only source (user-support discussions) which had the most response for the option ``more than average'' is an out-team human-centric source. The other sources which had the highest number of responses for the choice ``average'' stipulate that two are in-team human-centric (product backlog reviews, technical team discussions) and the other (user reviews) is out-team software-centric. Therefore, it can be said that the majority of the sources of NFR modifications are human-centric.
 
\textbf{NFR Deletion:} All three sources which give rise to NFR deletions on average are human-centric (in-team human-centric:product backlog review,technical team discussions; out-team human-centric:user-support discussions).
 
\textbf{NFR Bug Fix:} Out of the 5 sources which had ``average'' as the highest response, 2 are in-team software-centric (defect reports, individual developer's analysis), 2 are in-team human-centric (product backlog reviews, technical team discussions) and the other is out-team software-centric. Therefore, not as the other types of sources, NFR bug fixes originate mostly from software-centric sources.

\textbf{NFR Combination:} The sources which give rise to NFR combinations on average are mostly human-centric (in-team human-centric:product backlog reviews, technical team discussions; out-team human-centric: marketing team, user-support discussions) and the other two are software-centric(in-team software-centric: defect reports, out-team software-centric: user reviews).

\textbf{FR-NFR Combination:} Out of the 5 sources for which participants selected the choice ``average'' the most, 2 are in-team human-centric (product backlog reviews, technical team discussions), 2 are out-team human-centric (user reviews, user-support discussions) and the other is in-team software-centric (individual developer's detailed analysis). Therefore, majority of the sources of FR-NFR combinations are human-centric.

%% file: sections/where.tex
\emph{During iteration planning, daily standup, iteration review, iteration retrospective,} and \emph{after releasing the complete product} were the ceremonies/events where the RCs get originated as we found through our preliminary study. Along with these ceremonies/events, we offered the survey respondents the choice to provide any other ceremonies/events where RCs get originated. Similar to Section \ref{sec:source}, here we present highest number of responses given by the participants in the sub-sections. All results are given in Table \ref{tab:event}.

\begin{table*}[]
\caption{Ceremonies/Events Where Requirements Changes Originate (\includegraphics[scale=0.2]{figures/Scale.pdf}; FR: Functional Requirement; NFR: Non-Functional Requirement)}
\fontsize{14}{16}\selectfont
\label{tab:event}
\resizebox{\textwidth}{!}{%
\begin{tabular}{llllllllllllllll}
\hline
\multicolumn{1}{c}{\textbf{}}                & \multicolumn{3}{c}{\textbf{During Iteration Planning}}                                                  & \multicolumn{3}{c}{\textbf{During Daily Standup}}                                                       & \multicolumn{3}{c}{\textbf{During Iteration Review}}                                                    & \multicolumn{3}{c}{\textbf{\begin{tabular}[c]{@{}c@{}}During Iteration\\ Retrospective\end{tabular}}}   & \multicolumn{3}{c}{\textbf{\begin{tabular}[c]{@{}c@{}}After Releasing the\\ Complete Product\end{tabular}}} \\ \midrule
\multicolumn{16}{l}{\textbf{Functional Requirements Changes}}   \\ \midrule
FR Addition                                  & \multicolumn{3}{l}{\includegraphics[width=6cm,height=0.6cm]{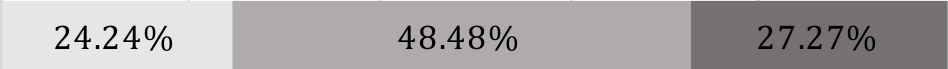}}                                                                           & \multicolumn{3}{l}{\includegraphics[width=6cm,height=0.6cm]{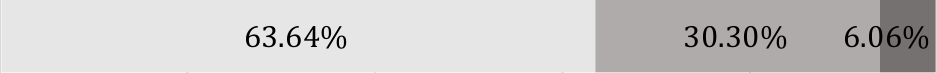}}                                                                           & \multicolumn{3}{l}{\includegraphics[width=6cm,height=0.6cm]{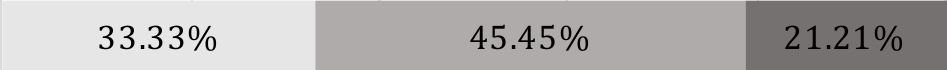}}                                                                                    & \multicolumn{3}{l}{\includegraphics[width=6cm,height=0.6cm]{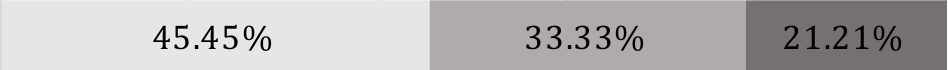}}                                                                                    & \multicolumn{3}{l}{\includegraphics[width=6cm,height=0.6cm]{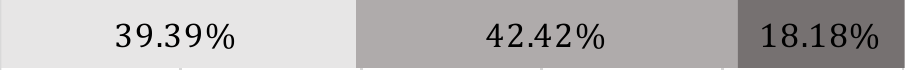}}                                                                                        \\
FR Modification                              & \multicolumn{3}{l}{\includegraphics[width=6cm,height=0.6cm]{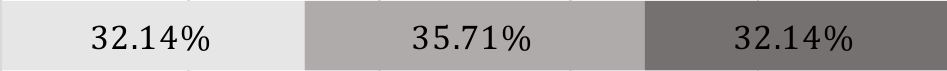}}                                                                           & \multicolumn{3}{l}{\includegraphics[width=6cm,height=0.6cm]{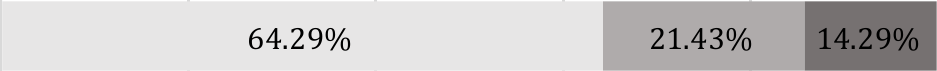}}                                                                           & \multicolumn{3}{l}{\includegraphics[width=6cm,height=0.6cm]{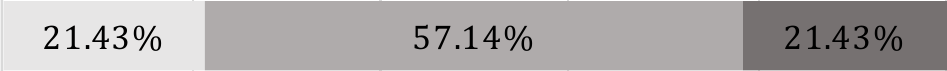}}                                                                                    & \multicolumn{3}{l}{\includegraphics[width=6cm,height=0.6cm]{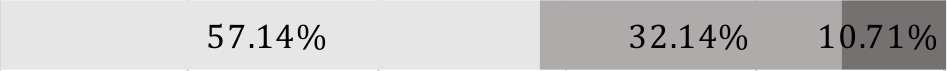}}                                                                                    & \multicolumn{3}{l}{\includegraphics[width=6cm,height=0.6cm]{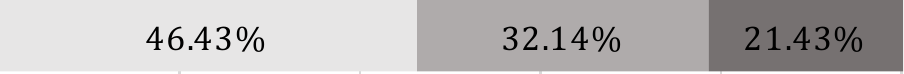}}                                                                                        \\
FR Deletion                                  & \multicolumn{3}{l}{\includegraphics[width=6cm,height=0.6cm]{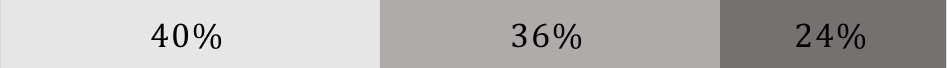}}                                                                           & \multicolumn{3}{l}{\includegraphics[width=6cm,height=0.6cm]{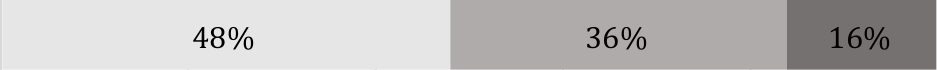}}                                                                           & \multicolumn{3}{l}{\includegraphics[width=6cm,height=0.6cm]{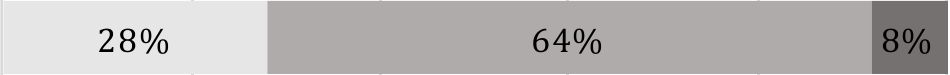}}                                                                                    & \multicolumn{3}{l}{\includegraphics[width=6cm,height=0.6cm]{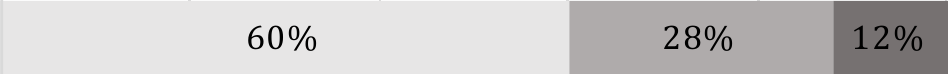}}                                                                                    & \multicolumn{3}{l}{\includegraphics[width=6cm,height=0.6cm]{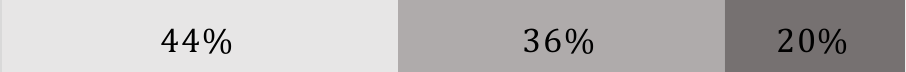}}                                                                                        \\
FR Bug Fix                                   & \multicolumn{3}{l}{\includegraphics[width=6cm,height=0.6cm]{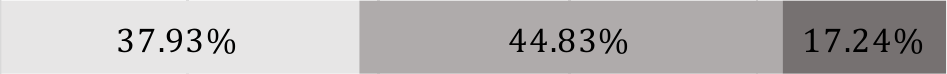}}                                                                           & \multicolumn{3}{l}{\includegraphics[width=6cm,height=0.6cm]{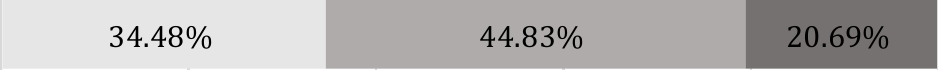}}                                                                           & \multicolumn{3}{l}{\includegraphics[width=6cm,height=0.6cm]{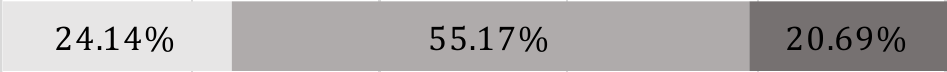}}                                                                                    & \multicolumn{3}{l}{\includegraphics[width=6cm,height=0.6cm]{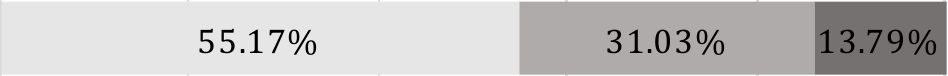}}                                                                           & \multicolumn{3}{l}{\includegraphics[width=6cm,height=0.6cm]{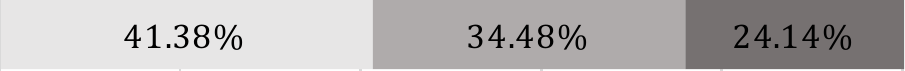}}                                                                                        \\
FR Combination                               & \multicolumn{3}{l}{\includegraphics[width=6cm,height=0.6cm]{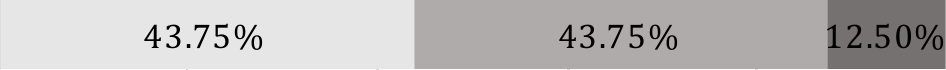}}                                                                           & \multicolumn{3}{l}{\includegraphics[width=6cm,height=0.6cm]{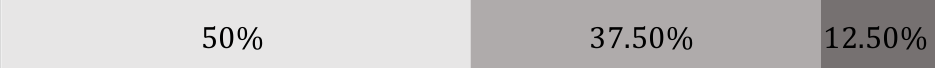}}                                                                           & \multicolumn{3}{l}{\includegraphics[width=6cm,height=0.6cm]{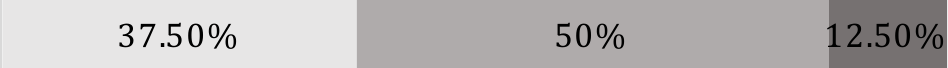}}                                                                                    & \multicolumn{3}{l}{\includegraphics[width=6cm,height=0.6cm]{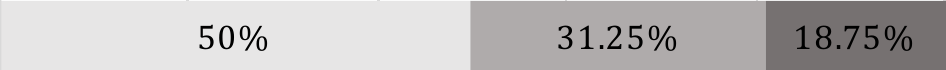}}                                                                                    & \multicolumn{3}{l}{\includegraphics[width=6cm,height=0.6cm]{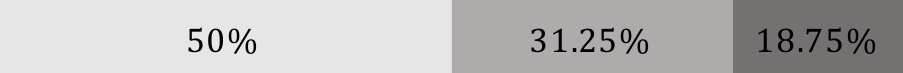}}                                                                                        \\ \midrule
\multicolumn{16}{l}{\textbf{Non-Functional Requirements Changes}} \\ \midrule
NFR Addition                                 & \multicolumn{3}{l}{\includegraphics[width=6cm,height=0.6cm]{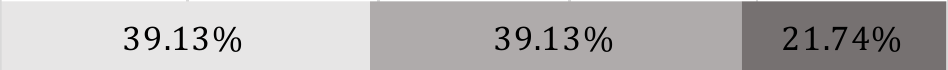}}                                                                                    & \multicolumn{3}{l}{\includegraphics[width=6cm,height=0.6cm]{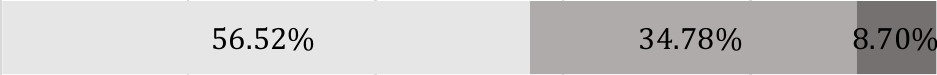}}                                                                                    & \multicolumn{3}{l}{\includegraphics[width=6cm,height=0.6cm]{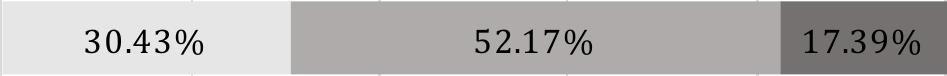}}                                                                          & \multicolumn{3}{l}{\includegraphics[width=6cm,height=0.6cm]{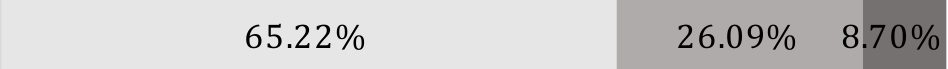}}                                                                           & \multicolumn{3}{l}{\includegraphics[width=6cm,height=0.6cm]{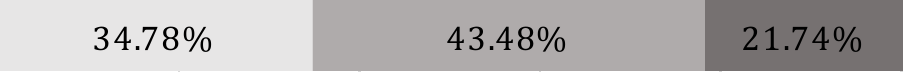}}                                                                               \\
NFR Modification                             & \multicolumn{3}{l}{\includegraphics[width=6cm,height=0.6cm]{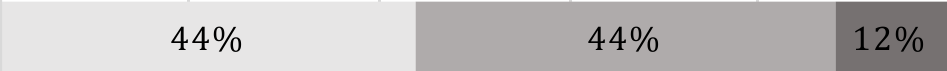}}                                                                                    & \multicolumn{3}{l}{\includegraphics[width=6cm,height=0.6cm]{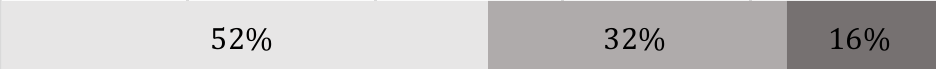}}                                                                                    & \multicolumn{3}{l}{\includegraphics[width=6cm,height=0.6cm]{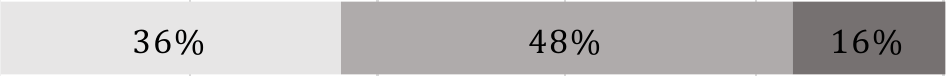}}                                                                           & \multicolumn{3}{l}{\includegraphics[width=6cm,height=0.6cm]{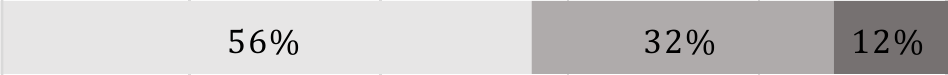}}                                                                           & \multicolumn{3}{l}{\includegraphics[width=6cm,height=0.6cm]{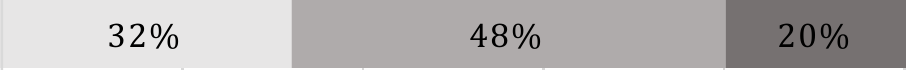}}                                                                                \\
NFR Deletion                                 & \multicolumn{3}{l}{\includegraphics[width=6cm,height=0.6cm]{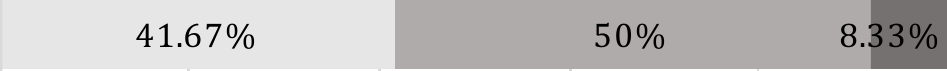}}                                                                                    & \multicolumn{3}{l}{\includegraphics[width=6cm,height=0.6cm]{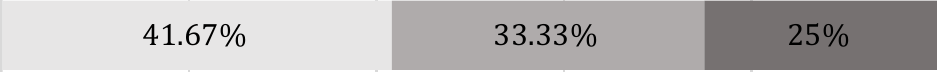}}                                                                                    & \multicolumn{3}{l}{\includegraphics[width=6cm,height=0.6cm]{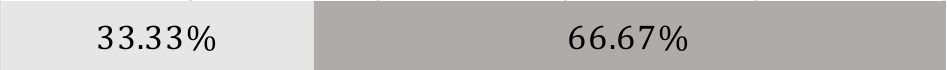}}                                                                           & \multicolumn{3}{l}{\includegraphics[width=6cm,height=0.6cm]{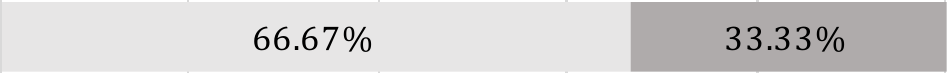}}                                                                           & \multicolumn{3}{l}{\includegraphics[width=6cm,height=0.6cm]{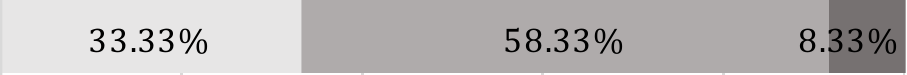}}                                                                                \\
NFR Bug Fix                                  & \multicolumn{3}{l}{\includegraphics[width=6cm,height=0.6cm]{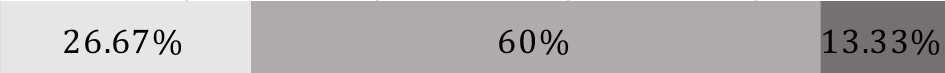}}                                                                                    & \multicolumn{3}{l}{\includegraphics[width=6cm,height=0.6cm]{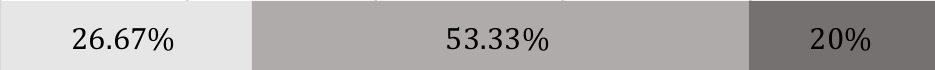}}                                                                           & \multicolumn{3}{l}{\includegraphics[width=6cm,height=0.6cm]{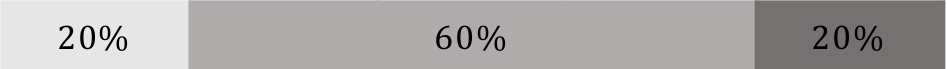}}                                                                           & \multicolumn{3}{l}{\includegraphics[width=6cm,height=0.6cm]{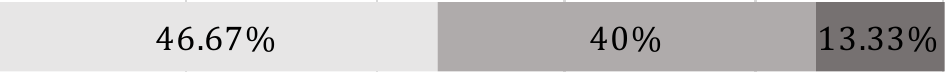}}                                                                           & \multicolumn{3}{l}{\includegraphics[width=6cm,height=0.6cm]{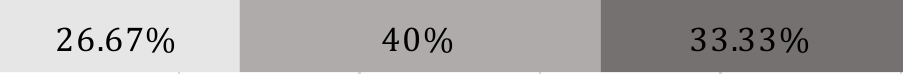}}                                                                                \\
NFR Combination                              & \multicolumn{3}{l}{\includegraphics[width=6cm,height=0.6cm]{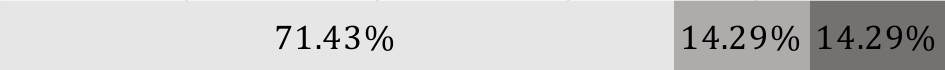}}                                                                                    & \multicolumn{3}{l}{\includegraphics[width=6cm,height=0.6cm]{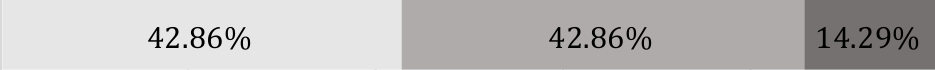}}                                                                           & \multicolumn{3}{l}{\includegraphics[width=6cm,height=0.6cm]{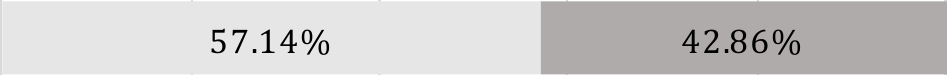}}                                                                           & \multicolumn{3}{l}{\includegraphics[width=6cm,height=0.6cm]{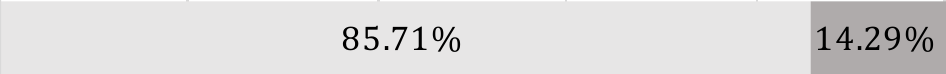}}                                                                           & \multicolumn{3}{l}{\includegraphics[width=6cm,height=0.6cm]{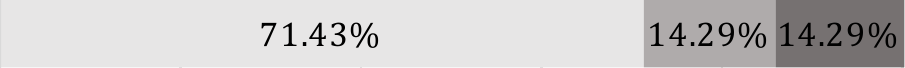}}                                                                                         \\ \midrule
FR-NFR Combination                           & \multicolumn{3}{l}{\includegraphics[width=6cm,height=0.6cm]{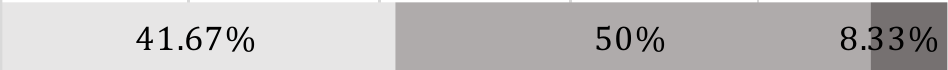}}                                                                           & \multicolumn{3}{l}{\includegraphics[width=6cm,height=0.6cm]{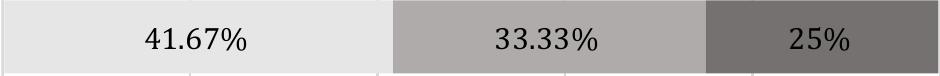}}                                                                           & \multicolumn{3}{l}{\includegraphics[width=6cm,height=0.6cm]{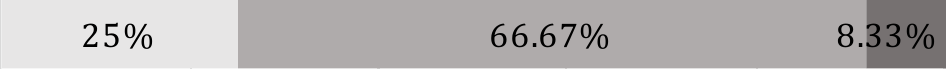}}                                                                           & \multicolumn{3}{l}{\includegraphics[width=6cm,height=0.6cm]{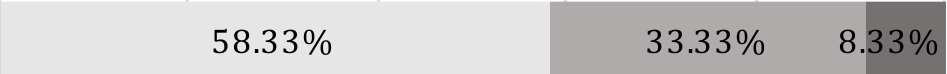}}                                                                           & \multicolumn{3}{l}{\includegraphics[width=6cm,height=0.6cm]{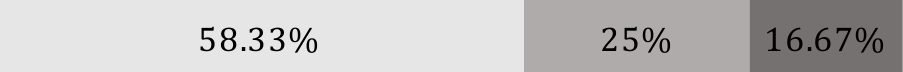}}                                                                                \\ \hline
\end{tabular}%
}
\end{table*}

\subsubsection{Ceremonies/Events Where Functional Requirements Changes Originate}

    \textbf{FR Addition:}
    Our findings show that FR additions originate on average during iteration planning (48.48\%), during iteration review (45.45\%), and after releasing the complete product (42.42\%). And it is unlikely for FR additions to occur during daily standup (63.64\%) and iteration retrospective (45.45\%) as most participants reported that the chance of FR additions' occurrence at those ceremonies/events is less than average.
    
    \textbf{FR Modification:}
     FR modifications originate on average during iteration planning (35.71\%) and during iteration review (57.14\%) whereas they originate less than average during daily standup (64.29\%), during iteration retrospective (57.14\%) and after releasing the complete product (46.43\%).
    
    \textbf{FR Deletion:}
    FR deletions originate during iteration planning (40\%), daily standup (48\%), iteration retrospective (60\%), and after releasing the complete product (44\%) is less than average.
    
    \textbf{FR Bug Fix:}
    FR bug fixes originate on average during iteration planning (44.83\%), daily standup (44.83\%), and during iteration review (55.17\%). There is less chance for the FR bug fixes to occur during iteration retrospective (55.17\%) and after releasing the complete product (41.38\%) as majority of the participants reported them as less than average.
    
    \textbf{FR Combination:}
    FR combinations also originate less than average during daily standup, iteration retrospective, and after releasing the complete product is less than average (50\% of resposes each) as reported by majority of participants. However, the same number of participants (43.75\% each) reported that FR combinations originate during iteration planning less than average and on an average of time. Therefore, it is vague to come to a conclusion about the frequency of FR combinations originating during iteration planning.
    
\subsubsection{Ceremonies/Events where Non-Functional Requirements Changes Originate}

    \textbf{NFR Addition:}
    NFR additions originate on average during iteration review (52.17\%) and after releasing the product (43.48\%). During daily standup (56.52\%) and iteration retrospective (65.22\%), NFR additions originate less than average as reported by majority of participants. However, the same number of participants (39.13\% each) reported that origination of NFR additions is less than average and on an average during iteration planning. Therefore, it is unclear if NFR additions originate during iteration planning on an average or less than average of time.
    
    \textbf{NFR Modification:}
    NFR modifications originate during iteration reviews (48\%) and after releasing the complete product (48\%) on an average. And also, during daily standup (52\%) and iteration retrospective (56\%) less than average as reported by the majority of the participants. Furthermore, it is less than average and on an average as reported by the most participants for the NFR modifications to originate during iteration planning (44\% each). Therefore, similar to NFR additions, we are unable to suggest if NFR modifications originate during iteration planning or not.
    
    \textbf{NFR Deletion:}
    NFR deletions originate during iteration planning (50\%), iteration review (66.67\%), and after releasing the complete product (58.33\%) on an average while NFR deletions originate during daily standup (41.67\%) and  during iteration retrospective (66.67\%) less than average as reported by majority of participants.

    \textbf{NFR Bug Fix:}
    NFR bug fixes originate on average during iteration planning (60\%), daily standup (53.33\%), iteration review (60\%), and after releasing the complete product (40\%). In addition, it is less likely for NFR bug fixes to originate during iteration retrospective 46.67\%) as reported by majority of participants.

    \textbf{NFR Combination:}
    The majority of responses were found as ``less than average'' for NFR combinations originating during iteration planning (71.43\%), iteration review (57.14\%), iteration retrospective (85.71\%), and after releasing the complete product (71.43\%). Same number of participants reported that NFR combinations originate less than average and on an average of time during daily standup (42.86\% each). Even though this is unclear, results indicate that it is less likely for NFR combinations to originate at any ceremonies/events except during daily standup. 

\subsubsection{Ceremonies/Events where Combinations of Functional and Non-Functional Requirements Originate}
    The chance of NFR-FR combinations to originate during iteration planning (50\%) and iteration review (66.67\%) is on average and during daily standup (41.67\%), iteration retrospective (58.33\%), and after releasing the complete product (58.33\%) is less than average.

\subsubsection{Other Events where Requirements Changes Originate}
A few other events where RCs originate were also reported by participants as given below.

\begin{itemize}
    \item \textbf{Quarterly planning meetings:} 
    When the duration of the entire product is  long,  such planning is required to be done. In this case, origination of RCs can be expected.

    \item \textbf{While coding:} 
    It is common to see the customers present if the team is onsite. In that case, the customer directly providing RCs during development can be expected, as reported by participant P28. Presence of a subject matter expert is possible at on-site or off-site. However, how the presence of user at the working premises and providing RCs is not clear. All these cases indicate that rather than following proper practices, free form of communicating requirements changes is done by various stakeholders:
    \begin{center}
        \textit{``The most common place is user/customer/SME directly working with the development team whilst coding the requirement.''  - P28 [Product Owner/Manager]}
    \end{center}

    \item \textbf{Workshops:} 
    When verbal communication (workshops) and interactions are prominent, there is a high chance for RCs to originate. As testers have high attention to detail when writing test cases, it is possible for RCs to originate: 
    \begin{center}
        \textit{``BA’s “unboxing” (workshopping) new User Stories with Dev Team; Testers trying to build test cases.'' - P24 [Tester]}
    \end{center}

    \item \textbf{Customer demos:}
    Usually customer demos are expected to occur at iteration reviews. However, our findings suggest that separate sessions for customer demos exist and RCs originate at customer demos.

\end{itemize}

%% file: sections/who.tex
\begin{table*}[]
\caption{Carriers of Requirements Changes (\includegraphics[scale=0.2]{figures/Scale.pdf}; FR: Functional Requirement; NFR: Non-Functional Requirement)}
\fontsize{14}{16}\selectfont
\label{tab:carrier}
\resizebox{\textwidth}{!}{%
\begin{tabular}{llcclcclcclcclcclcclcc}
\midrule
\multicolumn{1}{c}{\textbf{}} & \multicolumn{3}{c}{\textbf{Customer}}                           & \multicolumn{3}{c}{\textbf{Product Owner}}                      & \multicolumn{3}{c}{\textbf{Agile Coach}}                        & \multicolumn{3}{c}{\textbf{Developer}}                          & \multicolumn{3}{c}{\textbf{Other Agile Team Member}}            & \multicolumn{3}{c}{\textbf{Marketing Team}}                     & \multicolumn{3}{c}{\textbf{User-Support Team}}                  \\ \midrule
\multicolumn{22}{l}{\textbf{Functional Requirements Changes}}                                                                                                                                                                                                                                                                                                                                                                                                                                               \\ \midrule
FR Addition                   & \multicolumn{3}{l}{\includegraphics[width=5.2cm,height=0.6cm]{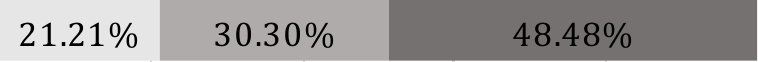}}                                            & \multicolumn{3}{l}{\includegraphics[width=5.2cm,height=0.6cm]{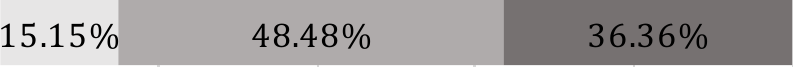}}                                            & \multicolumn{3}{l}{\includegraphics[width=5.2cm,height=0.6cm]{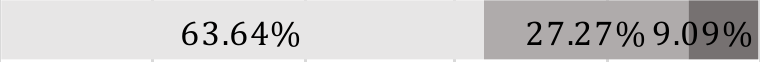}}                                   & \multicolumn{3}{l}{\includegraphics[width=5.2cm,height=0.6cm]{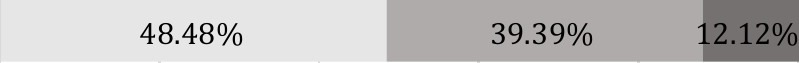}}                                   & \multicolumn{3}{l}{\includegraphics[width=5.2cm,height=0.6cm]{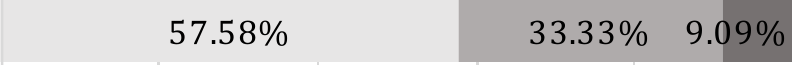}}                                   & \multicolumn{3}{l}{\includegraphics[width=5.2cm,height=0.6cm]{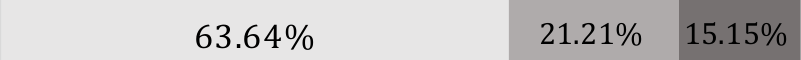}}                                   & \multicolumn{3}{l}{\includegraphics[width=5.2cm,height=0.6cm]{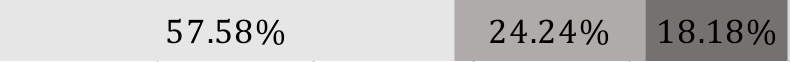}}                                   \\
FR Modification               & \multicolumn{3}{l}{\includegraphics[width=5.2cm,height=0.6cm]{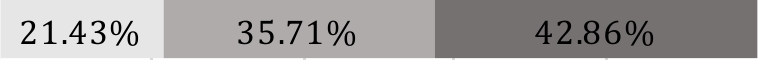}}                                            & \multicolumn{3}{l}{\includegraphics[width=5.2cm,height=0.6cm]{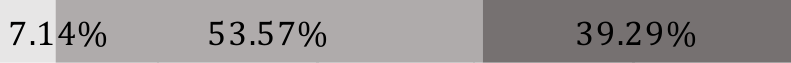}}                                            & \multicolumn{3}{l}{\includegraphics[width=5.2cm,height=0.6cm]{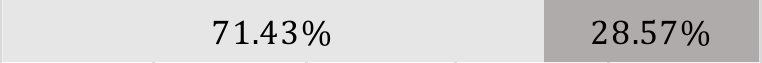}}                                   & \multicolumn{3}{l}{\includegraphics[width=5.2cm,height=0.6cm]{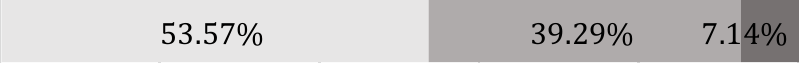}}                                   & \multicolumn{3}{l}{\includegraphics[width=5.2cm,height=0.6cm]{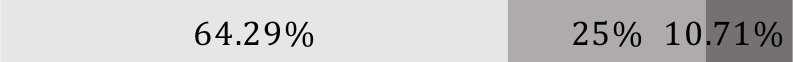}}                                   & \multicolumn{3}{l}{\includegraphics[width=5.2cm,height=0.6cm]{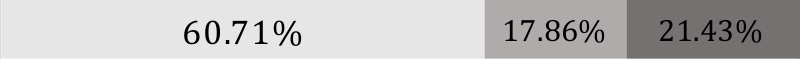}}                                   & \multicolumn{3}{l}{\includegraphics[width=5.2cm,height=0.6cm]{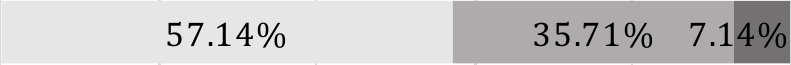}}                                   \\
FR Deletion                   & \multicolumn{3}{l}{\includegraphics[width=5.2cm,height=0.6cm]{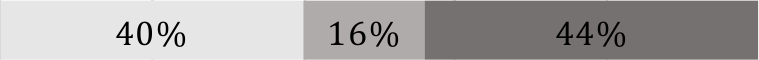}}                                            & \multicolumn{3}{l}{\includegraphics[width=5.2cm,height=0.6cm]{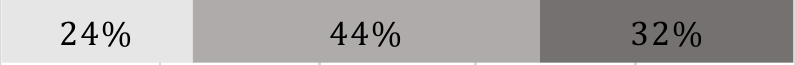}}                                            & \multicolumn{3}{l}{\includegraphics[width=5.2cm,height=0.6cm]{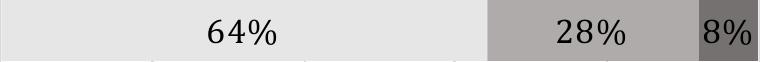}}                                   & \multicolumn{3}{l}{\includegraphics[width=5.2cm,height=0.6cm]{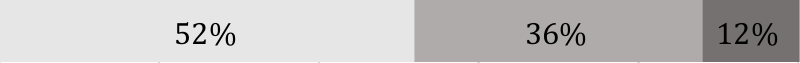}}                                   & \multicolumn{3}{l}{\includegraphics[width=5.2cm,height=0.6cm]{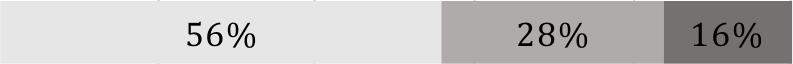}}                                   & \multicolumn{3}{l}{\includegraphics[width=5.2cm,height=0.6cm]{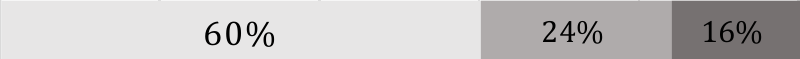}}                                   & \multicolumn{3}{l}{\includegraphics[width=5.2cm,height=0.6cm]{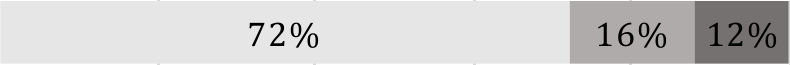}}                                   \\
FR Bug Fix                    & \multicolumn{3}{l}{\includegraphics[width=5.2cm,height=0.6cm]{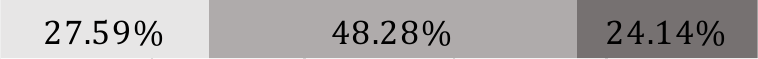}}                                            & \multicolumn{3}{l}{\includegraphics[width=5.2cm,height=0.6cm]{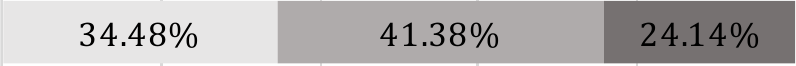}}                                            & \multicolumn{3}{l}{\includegraphics[width=5.2cm,height=0.6cm]{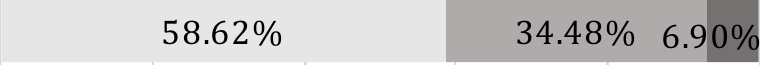}}                                   & \multicolumn{3}{l}{\includegraphics[width=5.2cm,height=0.6cm]{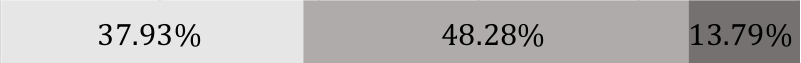}}                                            & \multicolumn{3}{l}{\includegraphics[width=5.2cm,height=0.6cm]{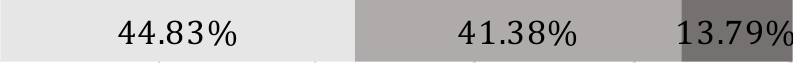}}                                   & \multicolumn{3}{l}{\includegraphics[width=5.2cm,height=0.6cm]{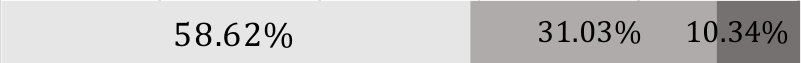}}                                   & \multicolumn{3}{l}{\includegraphics[width=5.2cm,height=0.6cm]{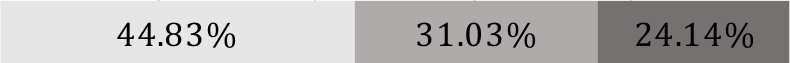}}                                   \\
FR Combination                & \multicolumn{3}{l}{\includegraphics[width=5.2cm,height=0.6cm]{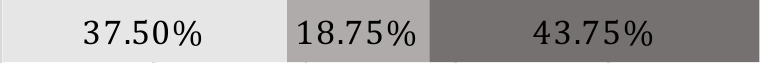}}                                            & \multicolumn{3}{l}{\includegraphics[width=5.2cm,height=0.6cm]{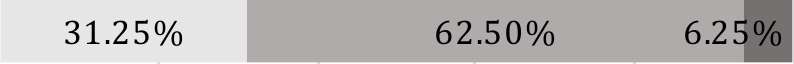}}                                            & \multicolumn{3}{l}{\includegraphics[width=5.2cm,height=0.6cm]{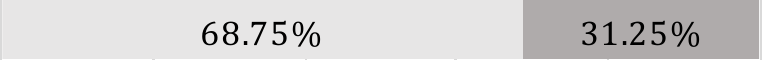}}                                   & \multicolumn{3}{l}{\includegraphics[width=5.2cm,height=0.6cm]{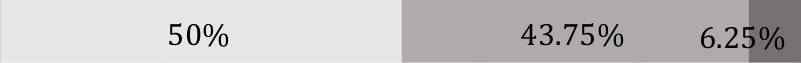}}                                   & \multicolumn{3}{l}{\includegraphics[width=5.2cm,height=0.6cm]{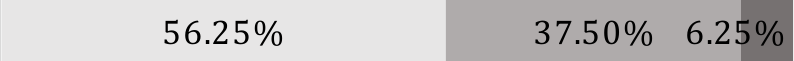}}                                   & \multicolumn{3}{l}{\includegraphics[width=5.2cm,height=0.6cm]{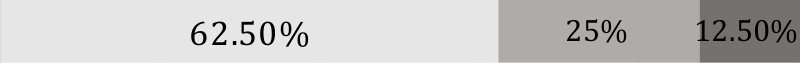}}                                   & \multicolumn{3}{l}{\includegraphics[width=5.2cm,height=0.6cm]{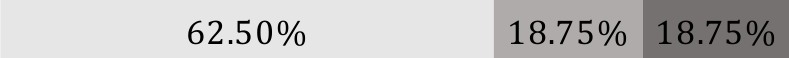}}                                   \\ \midrule
\multicolumn{22}{l}{\textbf{Non-Functional Requirements Changes}}                                                                                                                                                                                                                                                                                                                                                                                                                                           \\ \midrule
NFR Addition                  & \multicolumn{3}{l}{\includegraphics[width=5.2cm,height=0.6cm]{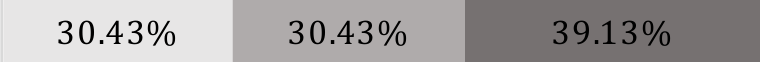}}                                            & \multicolumn{3}{l}{\includegraphics[width=5.2cm,height=0.6cm]{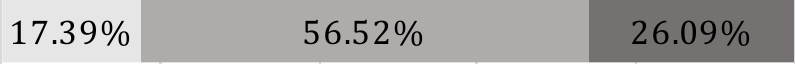}}                                            & \multicolumn{3}{l}{\includegraphics[width=5.2cm,height=0.6cm]{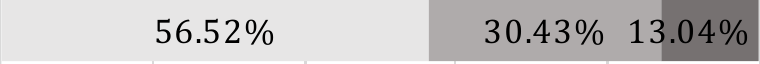}}                                   & \multicolumn{3}{l}{\includegraphics[width=5.2cm,height=0.6cm]{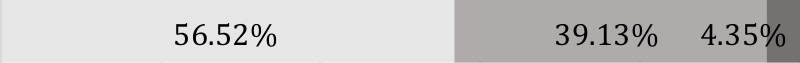}}                                   & \multicolumn{3}{l}{\includegraphics[width=5.2cm,height=0.6cm]{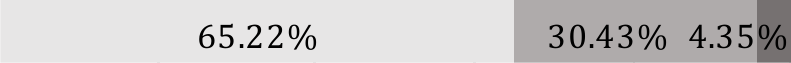}}                                   & \multicolumn{3}{l}{\includegraphics[width=5.2cm,height=0.6cm]{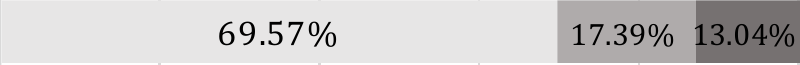}}                                   & \multicolumn{3}{l}{\includegraphics[width=5.2cm,height=0.6cm]{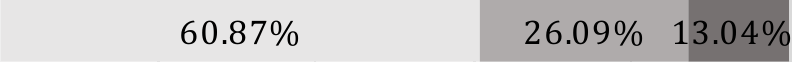}}                                   \\
NFR Modification              & \multicolumn{3}{l}{\includegraphics[width=5.2cm,height=0.6cm]{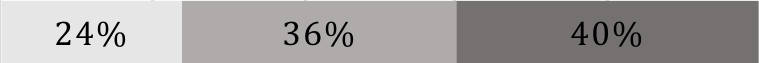}}                                            & \multicolumn{3}{l}{\includegraphics[width=5.2cm,height=0.6cm]{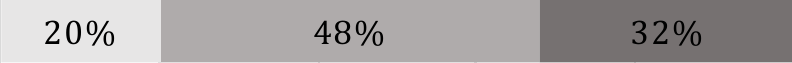}}                                            & \multicolumn{3}{l}{\includegraphics[width=5.2cm,height=0.6cm]{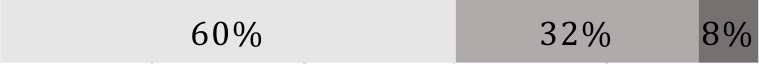}}                                   & \multicolumn{3}{l}{\includegraphics[width=5.2cm,height=0.6cm]{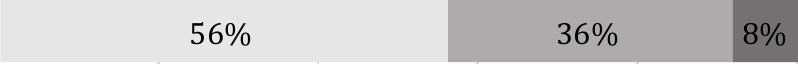}}                                   & \multicolumn{3}{l}{\includegraphics[width=5.2cm,height=0.6cm]{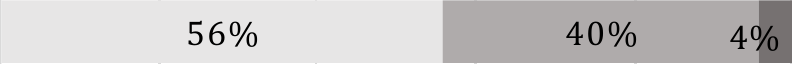}}                                   & \multicolumn{3}{l}{\includegraphics[width=5.2cm,height=0.6cm]{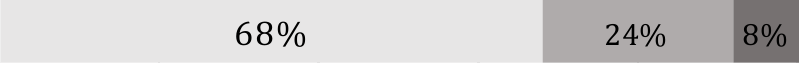}}                                   & \multicolumn{3}{l}{\includegraphics[width=5.2cm,height=0.6cm]{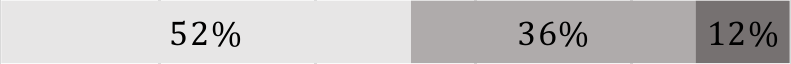}}                                   \\
NFR Deletion                  & \multicolumn{3}{l}{\includegraphics[width=5.2cm,height=0.6cm]{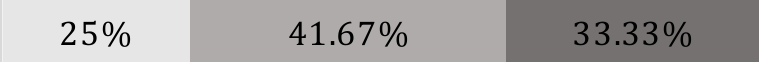}}                                            & \multicolumn{3}{l}{\includegraphics[width=5.2cm,height=0.6cm]{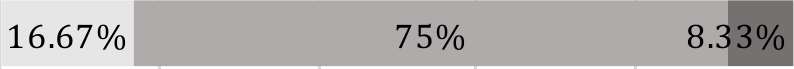}}                                            & \multicolumn{3}{l}{\includegraphics[width=5.2cm,height=0.6cm]{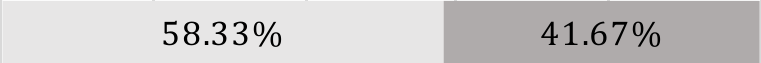}}                                   & \multicolumn{3}{l}{\includegraphics[width=5.2cm,height=0.6cm]{figures/who/developer/DevFRD.pdf}}                                   & \multicolumn{3}{l}{\includegraphics[width=5.2cm,height=0.6cm]{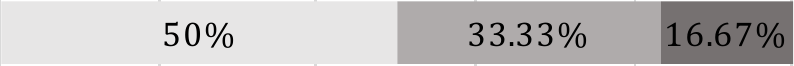}}                                   & \multicolumn{3}{l}{\includegraphics[width=5.2cm,height=0.6cm]{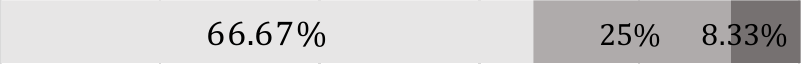}}                                   & \multicolumn{3}{l}{\includegraphics[width=5.2cm,height=0.6cm]{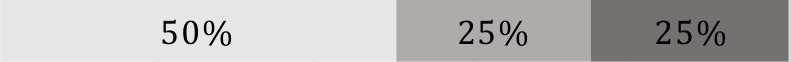}}                                   \\
NFR Bug Fix                   & \multicolumn{3}{l}{\includegraphics[width=5.2cm,height=0.6cm]{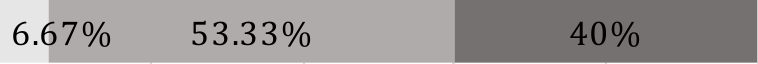}}                                            & \multicolumn{3}{l}{\includegraphics[width=5.2cm,height=0.6cm]{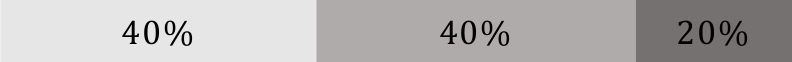}}                                  & \multicolumn{3}{l}{\includegraphics[width=5.2cm,height=0.6cm]{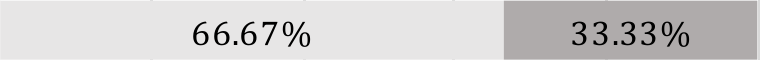}}                                   & \multicolumn{3}{l}{\includegraphics[width=5.2cm,height=0.6cm]{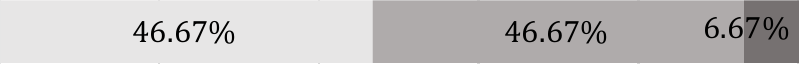}}                                   & \multicolumn{3}{l}{\includegraphics[width=5.2cm,height=0.6cm]{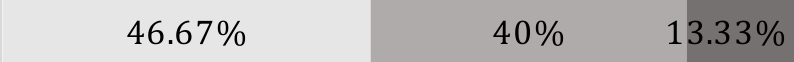}}                                   & \multicolumn{3}{l}{\includegraphics[width=5.2cm,height=0.6cm]{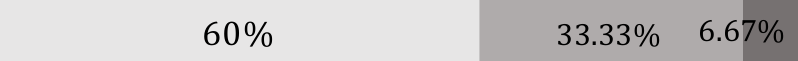}}                                   & \multicolumn{3}{l}{\includegraphics[width=5.2cm,height=0.6cm]{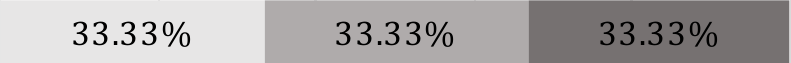}}                                   \\
NFR Combination               & \multicolumn{3}{l}{\includegraphics[width=5.2cm,height=0.6cm]{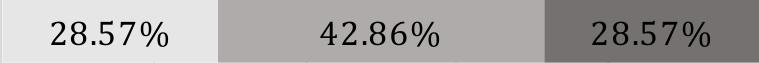}}                                            & \multicolumn{3}{l}{\includegraphics[width=5.2cm,height=0.6cm]{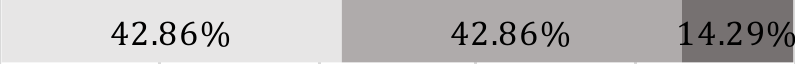}}                                   & \multicolumn{3}{l}{\includegraphics[width=5.2cm,height=0.6cm]{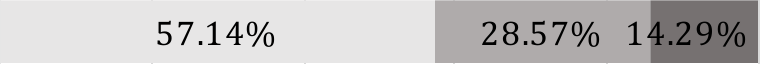}}                                   & \multicolumn{3}{l}{\includegraphics[width=5.2cm,height=0.6cm]{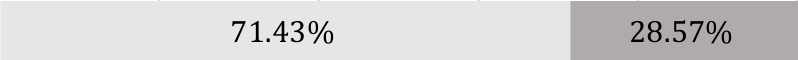}}                                   & \multicolumn{3}{l}{\includegraphics[width=5.2cm,height=0.6cm]{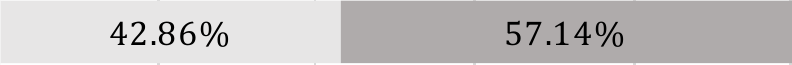}}                                            & \multicolumn{3}{l}{\includegraphics[width=5.2cm,height=0.6cm]{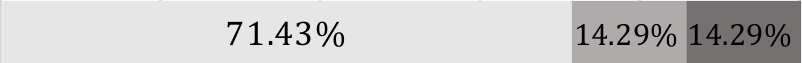}}                                   & \multicolumn{3}{l}{\includegraphics[width=5.2cm,height=0.6cm]{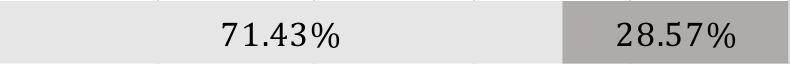}}                                   \\ \midrule
FR-NFR Combination            & \multicolumn{3}{l}{\includegraphics[width=5.2cm,height=0.6cm]{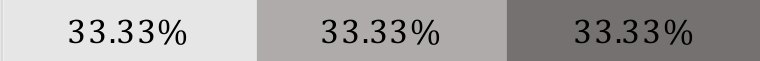}}                                   & \multicolumn{3}{l}{\includegraphics[width=5.2cm,height=0.6cm]{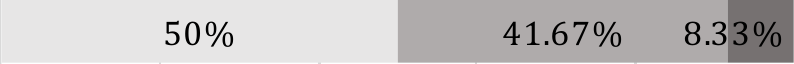}}                                   & \multicolumn{3}{l}{\includegraphics[width=5.2cm,height=0.6cm]{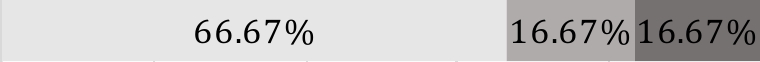}}                                   & \multicolumn{3}{l}{\includegraphics[width=5.2cm,height=0.6cm]{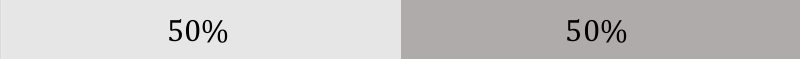}}                                   & \multicolumn{3}{l}{\includegraphics[width=5.2cm,height=0.6cm]{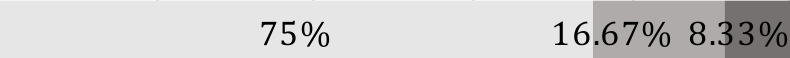}}                                   & \multicolumn{3}{l}{\includegraphics[width=5.2cm,height=0.6cm]{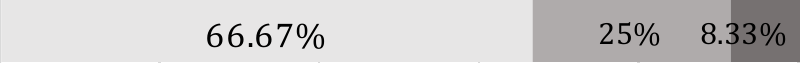}}                                   & \multicolumn{3}{l}{\includegraphics[width=5.2cm,height=0.6cm]{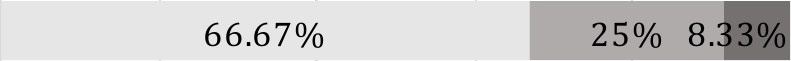}}                                   \\ \hline
\end{tabular}%
}
\end{table*}

We identified different requirements change 
``carriers" -- those who bring RCs to the team -- through our preliminary study. These carriers included the stakeholders: \textit{customer, product owner, agile coach, developer, other agile team member, marketing team,} and \textit{user support team}. By \textit{other agile team member}, we mean the members in the software development team other than developers such as business analysts, testers, and managers. The results are shown in Table \ref{tab:carrier}.

\subsubsection{Carriers who Bring Functional Requirements Changes}

\textbf{FR Addition:}
Customer (48.48\%) acts as the the carrier of FR additions more than average and product owner (48.48\%) plays the role of carrier on average as reported by the majority of participants. Other stakeholders (agile coach=63.64\%, developer=48.48\%, other agile team members=57.58\%, marketing team=63.64\%, user-support team=57.58\%) bring FR additions to the team less than average of time.

\textbf{FR Modification:}
Our findings show that customer (42.86\%) acts as the carrier of FR modifications more than average and product owner (53.57\%) being the carrier of FR modifications was on an average. Similar to the FR additions, agile coach (71.43\%), developer (53.57\%), other agile team members (64.29\%), marketing team (57.14\%), and user support team (57.14\%) bring FR modifications to the team less than average of time.

\textbf{FR Deletion:}
With regard to FR deletions, customer (42.86\%) acts as the carrier more than average, product owner (53.57\%) acts as the carrier in an average, and other stakeholders (agile coach=64\%, developer=52\%, other agile team members=56\%, marketing team=60\%, user-support team=72\%) act as the carrier less frequently as majority of participants reported that it is less than average.

\textbf{FR Bug Fix:}
Majority of responses were found for customer (48.28\%), developer (48.28\%), and product owner (41.38\%) acting as the carrier of FR bug fixes in an average whereas the other stakeholders (agile coach=48.62\%, other agile team members=44.83\%, marketing team=58.62\%, user-support team=44.83\%) play the role of carrier less than average of time.

\textbf{FR Combination:}
Customer (43.75\%) brings FR combinations to the team more than average of time, product owner (62.5\%) brings FR combinations on average, and other stakeholders (agile coach=68.75\%, developer=50\%, other agile team members=56.25\%, marketing team=62.5\%, user-support team=62.5\%) bring FR combinations to the team less than average of time.

\subsubsection{Carriers who Bring Non-Functional Requirements Changes}

\textbf{NFR Addition:}
Our findings confirm that customer (39.13\%) brings NFR additions to the team more than average, product owner (56.52\%) brings NFR additions to the team on an average and agile coach (56.52\%), developer (56.52\%), other agile team members (65.22\%), marketing team (69.57\%), and user-support team (60.87\%) bring NFR modifications to the team less than average.

\textbf{NFR Modification:}
According to the majority of responses,  customer (40\%) acts as the carrier of NFR modifications most of the time, product owner (48\%) acts as the carrier of NFR modifications on an average and agile coach (60\%), developer (56\%), other agile team members (56\%), marketing team (68\%), and user-support team (52\%) bring NFR modifications to the team less than average.

\textbf{NFR Deletion:}
Customer (41.67\%) and product owner (75\%) bring NFR deletions on an average to the team whereas other stakeholders (agile coach=58.33\%, developer=50\%, other agile team members=50\%, marketing team=66.67\%, user-support team=50\%) bring NFR deletions to the team less than average of time.

\textbf{NFR Bug Fix:}
Customer (53.33\%) brings NFR bug fixes to the team on average. Agile coach (66.67\%) brings NFR bug fixes to the team less than average as reported by the majority of participants. Developer (46.67\% each) and product owner (40\% each) bringing NFR bug fixes should be further studied as same number of responses were found for the options: less than average and more than average. Additionally, it is required to looked upon whether users-support team brings NFR bug fixes less than average, on average, or more than average as equal amount of responses (33.33\%) were found for the options.

\textbf{NFR Combination:}
Customer (42.86\%) and other agile team members (57.14\%) bring NFR combinations on the average. It is unclear if the product owner acts as the carrier of NFR combinations less than average or on average of time as same number of responses (42.86\% each) were found. Agile coach (57.14\%), developer (71.43\%), marketing team (71.43\%), and user-support team (71.43\%) bringing NFR combinations to the team is less than average of time.

\subsubsection{Carriers who Bring Combination of Functional and Non-Functional Requirements Changes}
The same number of responses (33.33\% each) were found for all three options: less than average, average, and more than average, customer being the carrier of FR-NFR combinations, it is required to look more for the most possible chance. Additionally, as the number of responses (50\% each) was divided between less than average and average for developer being the carrier, it is is required to further look into this as well. Other stakeholders (product owner=66.67\%, agile coach=66.67\%,  other agile team members=75\%, marketing team=66.67\%, user-support team=66.67\%) bring FR-NFR combinations less than average.

\subsubsection{Other Carriers who Bring Requirements Changes}
Apart from the carriers we provided for the participants to choose, carriers who fall under the category ``other agile team member'' as reported by the participants are listed below:

\begin{itemize}
    \item \textbf{Security team:} 
    In some cases, a security team exists in the project and they have the ability to identify NFR changes. Moreover, other stakeholders do not identify NFR changes as much as the security team, hinting that security teams are subject matter experts in NFRs whereas customer/product owner who brings other RCs often:
    \begin{center}
        \textit{``Our security team most often identifies NFR changes.'' - P37 [Product Owner]} 
    \end{center}
    
    \item \textbf{Business Analysts, Delivery Leads, Development Manager:} 
    These three roles are connected to interaction with external stakeholders such as customers and users. Therefore, it can be said that, due to their engagement with external stakeholders, they bring RCs at different circumstances such as critical requirements:
    \begin{center}
    \textit{``Other BAs, Delivery Leads \& Development Manager (especially on critical requirements).'' - P25 [Manager]} 
    \end{center}
\end{itemize}

%% file: sections/form.tex
\begin{figure}[b]
        \centering
        \includegraphics[width=\columnwidth]{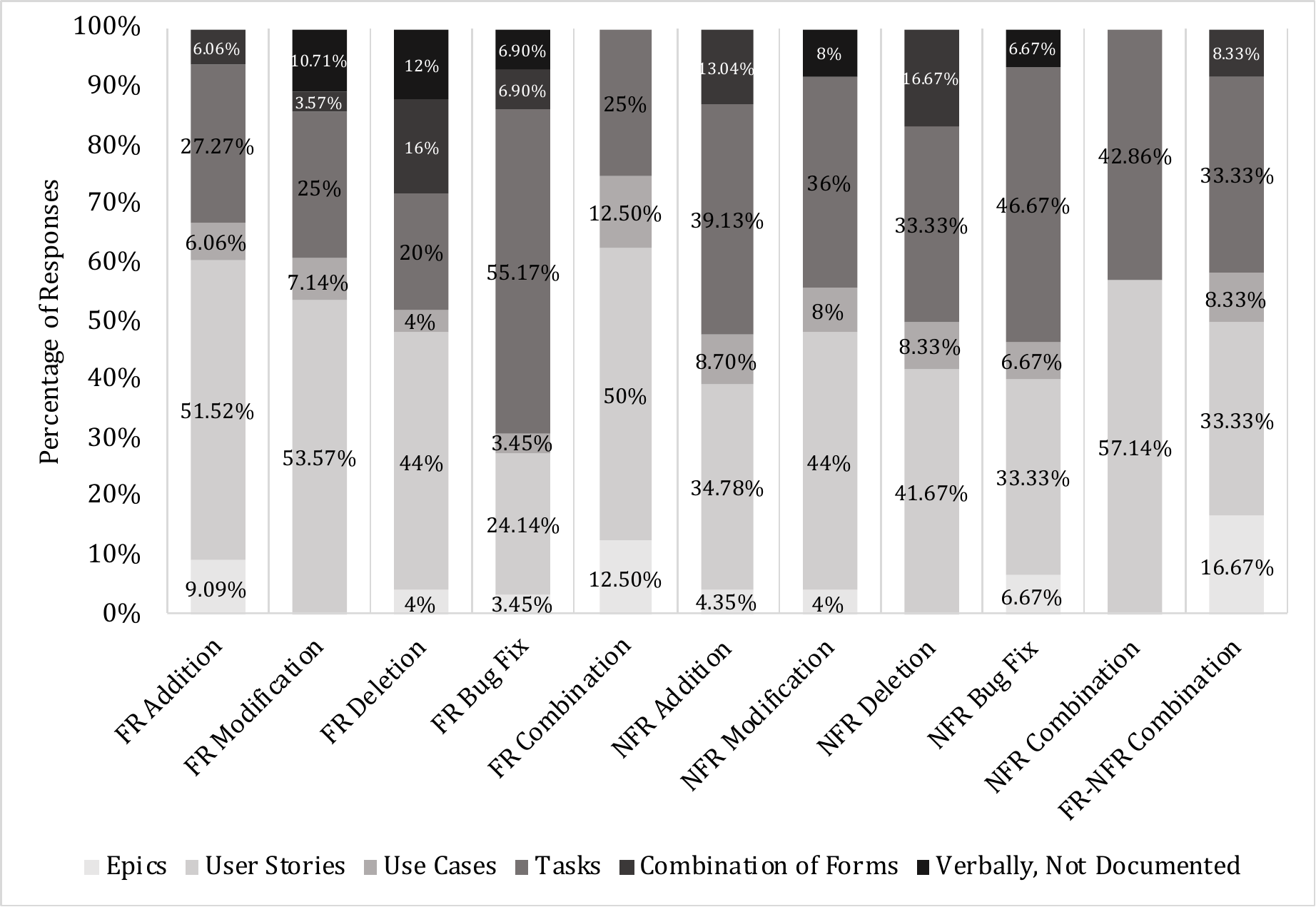}
        \caption{Documentation Forms of Requirements Changes}
        \label{fig:form}
    \end{figure}

Even though agile does not encourage use of detailed documentation \cite{Beck2001ManifestoDevelopment}, RCs are still documented for ease of use by the teams \cite{Madampe2020TowardsTeamsb, Paetsch2003}. RCs are documented in several forms such as \emph{epics, user stories, use cases}, and \emph{tasks}. Along with these forms, we provided \emph{combination of forms}, and \emph{verbally, not documented} for the participants to choose. As shown in Fig. \ref{fig:form}, participants chose the forms they use most commonly for each type of RC. By considering the results, it can be said that epics which are known to be a form that is widely used in agile environments are either not used or used at least by agile teams when documenting RCs. User stories and tasks were most preferable by the participants to document RCs. 

Surprisingly, FR modifications, FR deletions, FR bug fixes, FR combinations, NFR modifications, and NFR bug fixes are verbally provided to the teams but are not documented. Not documenting the RCs may lead to serious concerns directly in project schedule, cost, and also may impact the quality of the software. Use cases which are a form known to be used in traditional software development methods are still in use in agile environments as well, but in a lesser amount as reported by the participants. Additionally, combination of forms was selected by the participants in a lesser number as a choice for documenting RCs. In the following sub-sections, we report the most common responses given by the participants for each RC type.

\subsubsection{Documentation Forms of Functional Requirements Changes}
\emph{User stories} was reported as the most common form of documenting FR additions (51.52\%), FR modifications (53.57\%), FR deletions (44\%) and FR combinations (50\%). Majority of the participants (55.17\%) reported FR bug fixes are documented in the form of \textit{tasks}.

\subsubsection{Documentation Forms of Non-Functional Requirements Changes}
Majority of the participants reported that they use \textit{user stories} to document NFR modifications (44\%), NFR deletions (41.67\%), and NFR combinations (57.14\%); and tasks to document NFR additions (39.13\%) and NFR bug fixes (46.67\%).

\subsubsection{Documentation Forms of Combination of Functional and Non-Functional Requirements Changes}

\textit{User stories} and \textit{tasks} were equally reported (33.33\%) as the highest responded choice for documenting FR-NFR combination RCs. 

%% file: sections/challenge.tex
We considered \emph{complexity, cascading impact, size of RC, effort required, definition, priority,} and \emph{access to customer} as the metrics to measure the challenging nature of any given RC. According to Boehm \cite{Boehm1984SoftwareEconomics}, \textit{complexity} is one of the important drivers in software cost. We combined requirements dependability and change conflicts with existing requirements which are considered as challenges in RC management in general \cite{Anwer2019AResults} as \textit{cascading impact}. Furthermore, we derived \emph{access to customer} from our interview-based study findings and also adapted from Anwer et al.'s work \cite{Anwer2019AResults} along with \emph{prioritization} as communication is considered as a challenge in RC management in global software development and prioritization is a challenge in RC management in general, respectively. Other metrics: size of RC, effort required, and definition were hypothesized. Complexity, cascading impact, effort required, and priority followed the dimensions low, medium, high.

We used small, medium, and large as the dimensions for \textit{size of RC}. The dimensions imprecise or unclear, doesn't matter, and precise and clear were used for the metric \textit{definition}. Difficult or irregular, doesn't matter, and easy and regular were used for the metric \textit{access to customer} as shown in Table \ref{tab:challenge}.

\begin{table}
\caption{Requirements Change Challenging Metrics (RC: Requirements Change)}
\label{tab:challenge}
\resizebox{\columnwidth}{!}{%
\begin{tabular}{@{}llll@{}}
\toprule
\textbf{Metric}  & \multicolumn{3}{c}{\textbf{Dimension}}                                                                \\ \midrule
Complexity         & Low (4\%)                              & Medium (22\%)         & \textbf{High (74\%)}    \\
Cascading Impact   & Low (4\%)                              & Medium (34\%)         & \textbf{High (62\%)}    \\
Size of RC         & Small (13\%)                           & Medium (30\%)         & \textbf{Large (57\%)}   \\
Effort Required    & Low (4\%)                              & Medium (32\%)         & \textbf{High (64\%)}    \\
Definition         & \textbf{Imprecise or Unclear (71\%)}   & Doesn't Matter (22\%) & Precise and Clear (7\%) \\
Priority           & Low (7\%)                              & Medium (20\%)         & \textbf{High (73\%)}    \\
Access to Customer & \textbf{Difficult or Irregular (65\%)} & Doesn't Matter (21\%) & Easy and Regular (14\%) \\ \bottomrule
\end{tabular}
}
\end{table}

Taking the top most responded results by the participants as given in Table \ref{tab:challenge} into consideration, an RC is seen as challenging when its,
\begin{itemize}
    \item \textbf{complexity} is \textbf{high} and/or;
    \item \textbf{cascading impact} is \textbf{high} and/or;
    \item \textbf{size} is \textbf{large} and/or;
    \item \textbf{effort required} is \textbf{high} and/or;
    \item \textbf{definition} is \textbf{imprecise or unclear} and/or;
    \item \textbf{priority} is \textbf{high} and/or;
    \item \textbf{access to customer} is \textbf{difficult or irregular}.
\end{itemize}

\begin{figure}[b]
    \centering
    \includegraphics[width=\columnwidth]{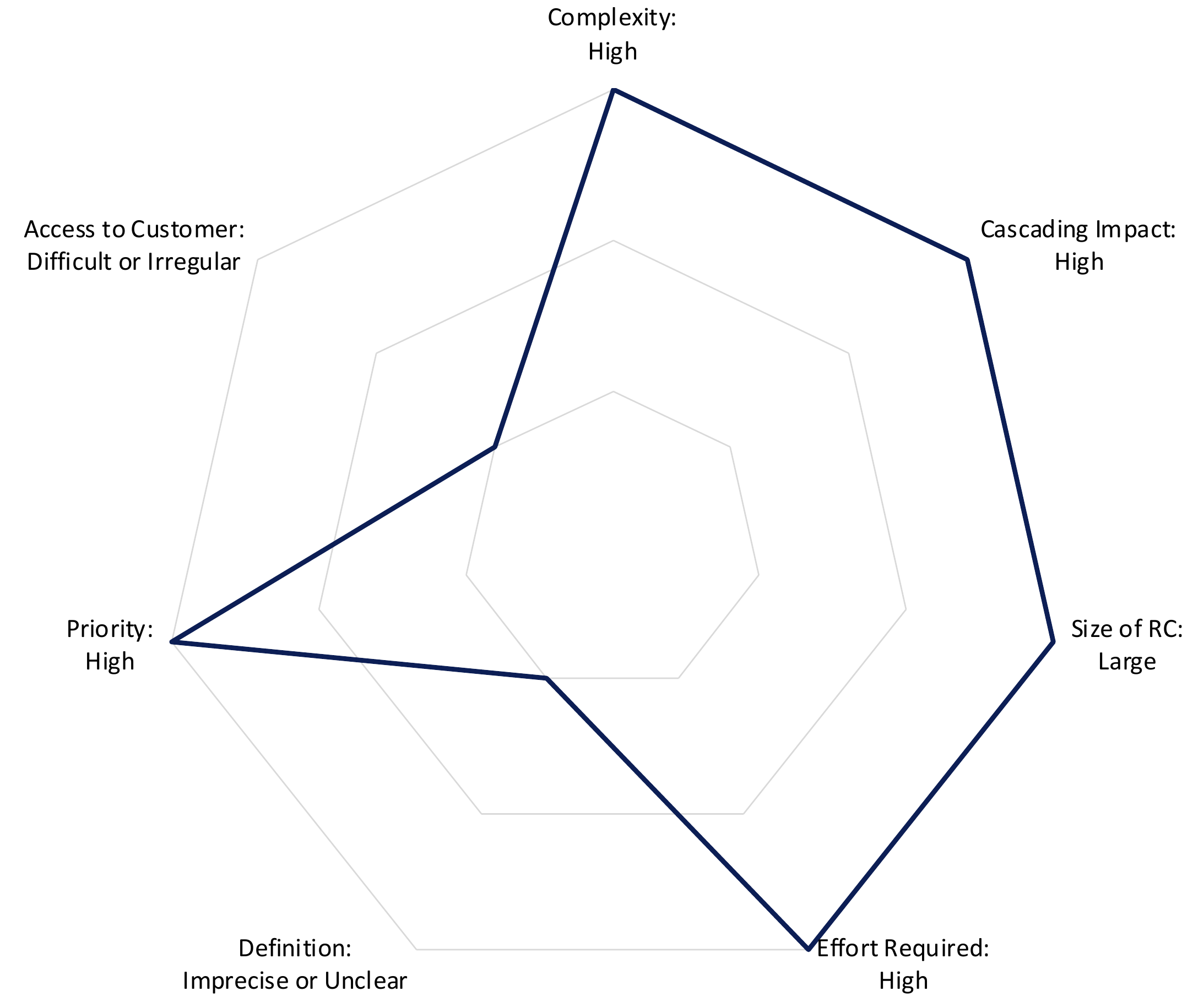}
    \caption{Conceptual Framework: Metrics to Measure the Challenging Nature of Requirements Changes in Agile Environments}
    \label{fig:challenge}
\end{figure}

We constructed a conceptual framework using these metrics to measure the challenging nature of  RCs in agile environments, an example is given in Fig. \ref{fig:challenge}. This framework can be used in assisting development teams with self-assignment \cite{Masood2020HowStudy}.

\subsubsection{Factors Intensifying the Challenging Nature of Requirements Changes}
Apart from complexity, cascading impact, size, effort required, definition, priority, and access to customer given in the conceptual framework, our further analysis resulted in identifying the challenging nature of RCs become intensified due to \textit{forced cross-functionality, organizational needs, complex development,} and \textit{time-boxed nature of agile} as summarized in Table \ref{tab:other_challenges}. However, improvements can be done within the team so as to avoid such discomforts.
     
     \begin{table}[]
\caption{Preliminary Suggestions on Factors Intensifying the Challenging Nature of Requirements Changes}
\label{tab:other_challenges}
\resizebox{\columnwidth}{!}{%
\begin{tabular}{@{}l@{}}
\toprule
\textbf{Requirements Changes Become Challenging Further Due to} \\ \midrule
\rowcolor[HTML]{C0C0C0} 
\textbf{Forced Cross-Functionality}                                                   \\
Business analysts forcing to complete the development           \\
Lack of analysis by developers                                  \\
Developers being disengaged from thinking                       \\
\rowcolor[HTML]{C0C0C0} 
\textbf{Organizational Needs       }                                     \\
Company policies                                                \\
Regulatory                                                      \\
Legal aspects                                                   \\
Service level agreements                                        \\
Business rules                                                  \\
\rowcolor[HTML]{C0C0C0} 
\rowcolor[HTML]{C0C0C0} 
\textbf{Time-boxed nature of agile }                                     \\ \bottomrule
\end{tabular}%
}
\end{table}

    Business analysts forcing to complete the development, lack of analysis by developers, and developers being disengaged from thinking sums up to forced cross-functionality:
    \begin{center}
    \textit{``..disengaged from thinking \& expecting others to do the thinking \& exploring of expected business value (“I just want to write code, man!”)'' - P25 [Manager]}
    \end{center}

Organizational needs such as company policies, regulatory, legal aspects, service level agreements, and business rules, which are not under the control of the agile team make the RCs more challenging to the team. Furthermore, we found that time-boxed nature of agile also can worsen the situation. However, these concepts are not matured enough as these were mentioned only by one or two participants. Therefore, it is required to study them more on this in the future.

%% file: sections/agilehelps.tex
\subsubsection{Requirements Changes as a Percentage of Total Work Items}

\begin{figure}[b]
        \centering
        \includegraphics[width=\columnwidth]{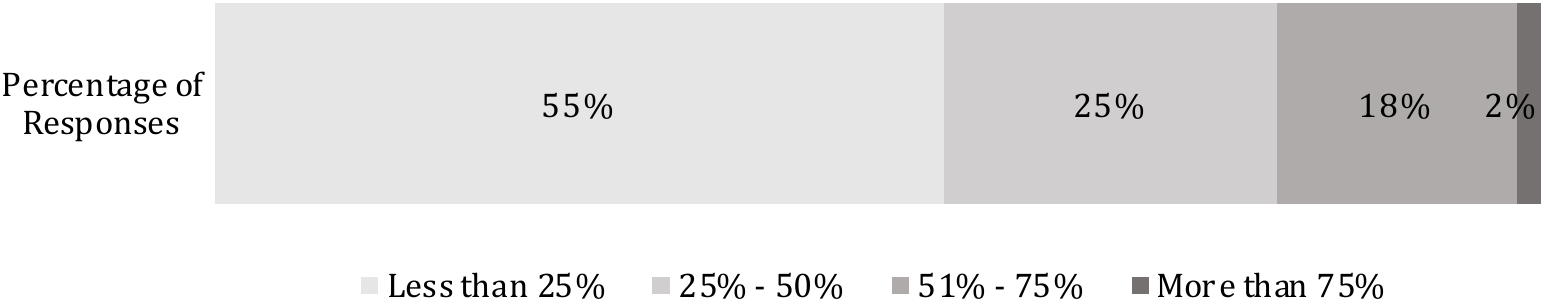}
        \caption{Requirements Changes as a Percentage of Total Work Items}
        \label{fig:percentage_RC}
    \end{figure}

It was surprising that 55\% of the participants responded that  less than 25\% of the total work items in the project were RCs (Fig. \ref{fig:percentage_RC}). 25\% of the participants reported that RCs are between 25\% - 50\% of the total work items. In addition, 18\% and 2\% of the participants reported that RCs are between 51\% - 75\% and more than 75\% of the work items respectively. If agile is all about embracing changes, why only 1/4 of the total work items were RCs and why rest were fixed requirements? This gives rise to three questions to look into in the future,
\begin{itemize}
    \item Do the teams practice hybrid versions of agile and traditional software development even though they claim that what they practice is agile?
    \item Why do teams practice agile if 75\% of the total work items are fixed?
    \item If the remaining work items are neither RCs nor fixed, what are they?
\end{itemize}
These questions need to be investigated in future studies.

\subsubsection{Why Agile?}
We asked the participants why do they use agile irrespective of their response on agile's ability to respond to RCs. As given in Table \ref{tab:why_agile}, we found several reasons which we categorized as human-centric and software-centric reasons.

\textbf{Human-centric positive reasons:}
Even though a participant mentioned that agile helps to respond to RCs, several other reasons were provided by other participants. These reasons include, being able to deliver the product in a shorter time period (N=13), surface any issues quickly (N=2), receive constant customer feedback (N=9), increase productivity (N=17), split the tasks into granular tasks (N=1), active and close collaboration (N=2), continuous learning opportunities for the team (N=1), experiment with the tasks (N=1), have a team empowered with autonomy (N=1), and improve communication (N=1). However, the top three reasons for practicing agile as mentioned by \textit{State of Agile Survey} \cite{202014thAgile} are accelerating software delivery, enhance ability to manage changing priorities, and increase productivity whereas almost half of our participants disagreed with the fact that it enhances ability to manage changing priorities (N=19).

\textbf{Software-centric positive reasons:}
Agile helps to remove unnecessary modifications (N=1), improve the product continuously (N=2), enhance the visibility (N=1), and improve the product quality (N=1) as reported by the participants.
We also found that, some participants (N=8) practice agile due to either organizational needs or customer needs. That is, organization or customer require the project to be run in an agile environment.

\subsubsection{Drawbacks of Agile}
The benefits of using agile are discussed much more in research literature than its drawbacks. Through our survey, we found that iteration pressure which is rooted due to working under sustainable pressure to deliver customer value \cite{Babb2014EmbeddingDevelopment} causes stress within team members (N=1). This is a major concern as agile is said to be a human-centric software development methodology. If such a methodology causes harm to the humans involved in it, then its human-centric property must be in question. Furthermore, we found that agile does not always allow teams to meet their customer's expectations. In such instances, practicing agile is de-prioritized:

\begin{center}
    \textit{``Agile becomes something else when it comes to client's needs and at that point, no one cares about Agile'' - P27 [Developer]}
\end{center}


\begin{table}[]
\caption{Why Practicing Agile?}
\label{tab:why_agile}
\resizebox{\columnwidth}{!}{%
\begin{tabular}{ll}
\toprule
\multicolumn{1}{c}{\textbf{Reason for Practicing Agile}} & \multicolumn{1}{c}{\textbf{Category}}      \\ \midrule
Responding to requirements changes              & \multirow{11}{*}{Human-Centric}   \\
Deliver in a shorter time period                &                                   \\
Surfacing any issues quickly                    &                                   \\
Constant feedback from customer                 &                                   \\
Increases productivity                          &                                   \\
Ability to split tasks into granular tasks      &                                   \\
Active and close collaboration                  &                                   \\
Continuous learning for the team                &                                   \\
Supports experimentation                        &                                   \\
Empowers team autonomy                          &                                   \\
Improves communication                          &                                   \\ \midrule
Helps to remove unnecessary modifications       & \multirow{4}{*}{Software-Centric} \\
Continuous product improvement                  &                                   \\
Enhances visibility                             &                                   \\
Improves product quality                        &                                   \\ \midrule
Doing agile for organizational/customer needs   & General                              \\ \bottomrule
\end{tabular}%
}
\end{table}
    
\subsubsection{Using Strategies to Minimize Requirements Changes}
\begin{figure}[b]
    \centering
\includegraphics[width=\columnwidth]{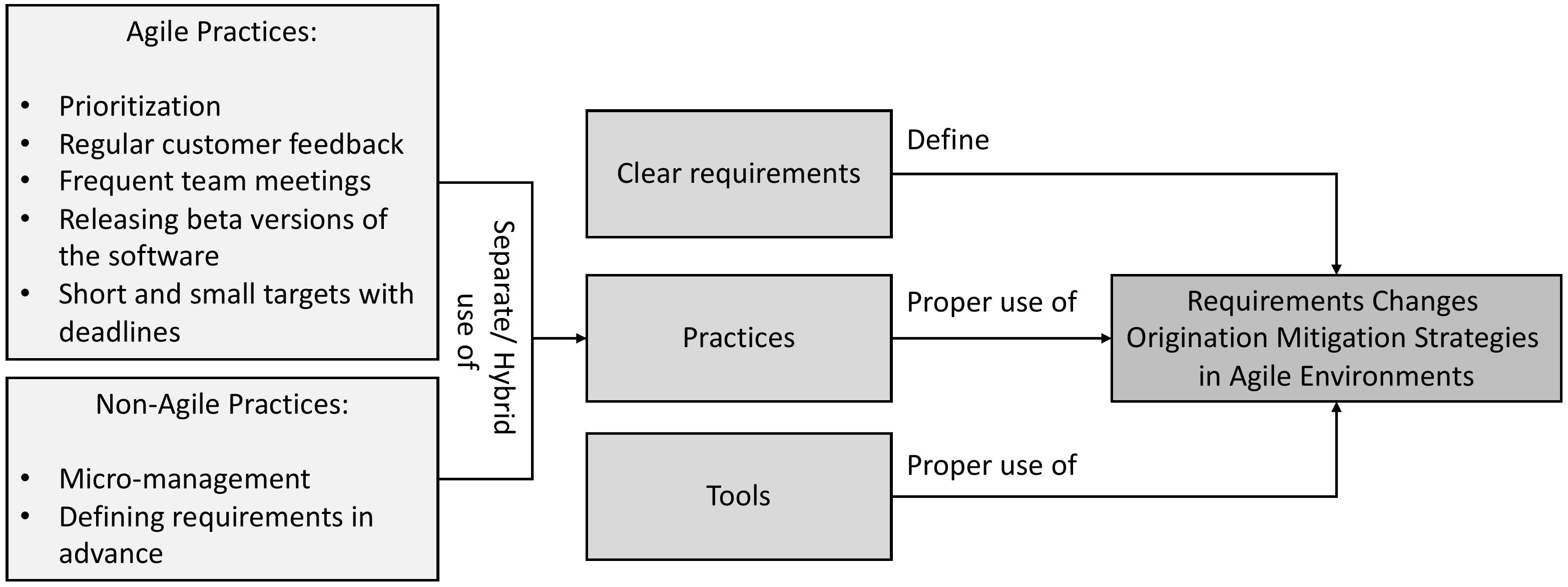}
    \caption{Requirements Change Minimization Strategies}
    \label{fig:mitigation}
\end{figure}

We asked for the reason why less requirements changes were received from the participants who stated this. By analysing the data, we found that, they try to mitigate the origination of RCs and use different strategies as shown in Fig. \ref{fig:mitigation}. Defining requirements clearly and using practices and tools properly are the strategies we found.

\textbf{Defining requirements clearly:}
As reported by the participants, if RCs are welcomed, changes have to be made in the epics and/or user stories; which is not preferred by the teams. Moreover, changing these \textit{interrupts the team focus} and \textit{impacts the software delivery}. Therefore, it is needed to define the requirements more clearly:
\begin{center}
    \textit{``In my projects, clear requirements of the project is key..'' - P20 [Product Owner]}
\end{center}

\textbf{Proper use of practices:}
Teams mix and match agile (prioritization, regular customer feedback, frequent team meeting, releasing beta versions of the software, short and small targets with deadlines) and non-agile (micro-management, defining requirements in advance) practices to block RCs. Especially, non-agile practices which are not recommended by agile are in use as reported by the participants. According to participant P16, they achieve the accuracy with the help of \textit{micro-management}:

\begin{center}
    \textit{``Accuracy of project deliverables is because of micro level supervision'' - P16[Agile Coach/Scrum Master, Product Owner, Developer, Tech Lead] 
}
\end{center}

\textbf{Proper use of tools:}
As stated by the participants, using tools (E.g.: \textit{JIRA}) meaningfully, helps them to stop the RCs from originating. This is also one of the reasons why they prefer to have their projects use agile methods.

\subsubsection{Facing Difficulties After Accepting Requirements Changes}
We found that teams face difficulties after they accept RCs. As reported by the participants, the changes in epics/user stories due to large number of RCs make it difficult for them to work. In addition, they find it difficult to adjust their work when the RCs are received in the middle of an iteration and expected to develop the RCs in the current iteration in which they already have user stories to work on. Furthermore, participants reported that being difficult to adjust in the middle of the iteration results in impacting the product delivery and interrupting the team focus:

\begin{center}
    \textit{``Too many changes may not really work well with the agile framework because that would cause epic/use stories to change. Changing the user stories in the middle of the sprint may interrupt team focus and impact the delivery..'' - P20 [Product Owner]}
\end{center}

This suggests that even though the teams are practicing agile, they are expected to reshape the agile practices in order to fit accordingly to the respective circumstance. However, according to the Scrum Guide \cite{2015TheGuide}, no changes should be made that endanger the sprint goal.

\subsubsection{Using Techniques to Handle Requirements Changes}
\begin{figure}
    \centering
    \includegraphics[width=\columnwidth]{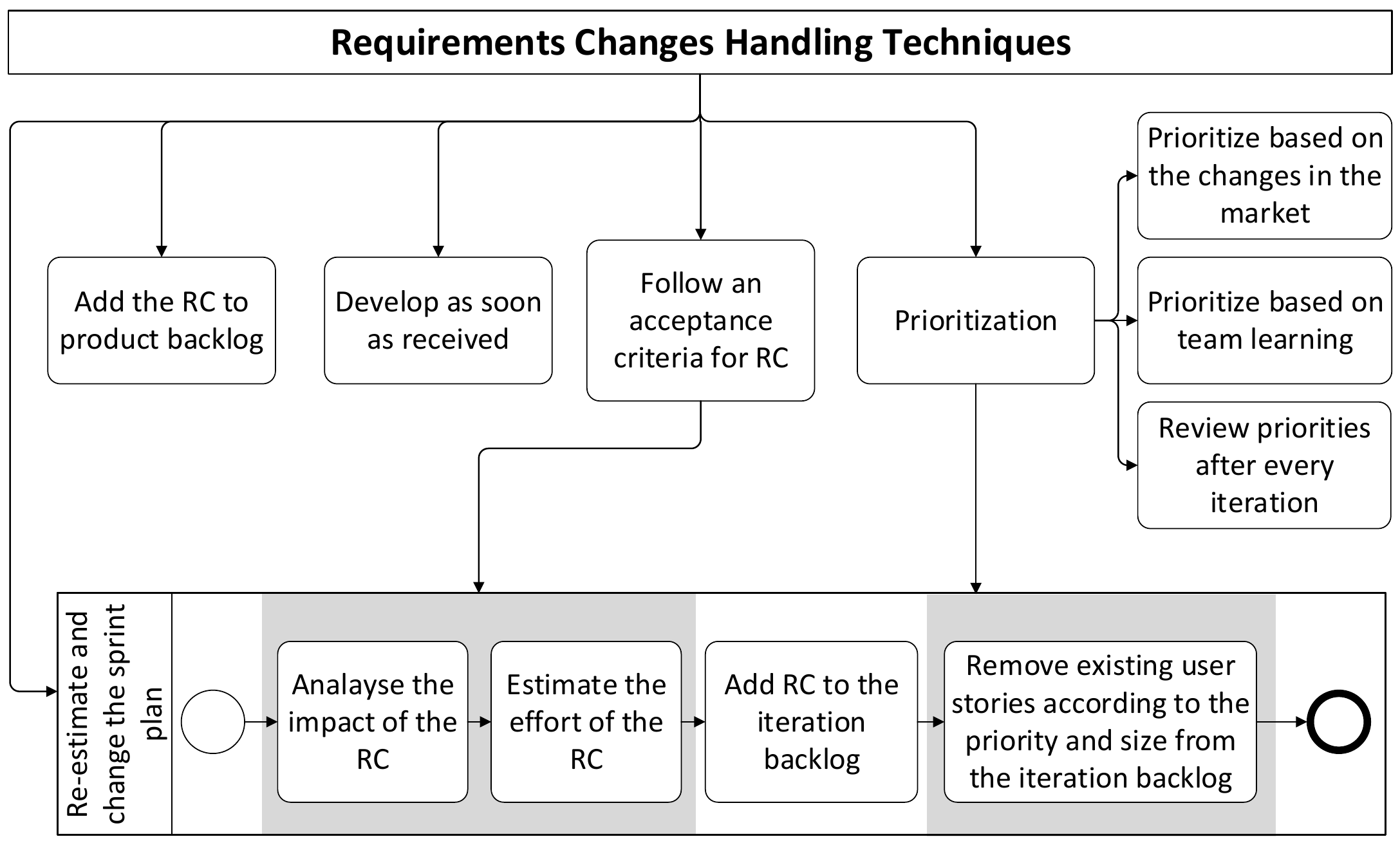}
    \caption{Requirements Changes Handling Techniques (RC: Requirements Change)}
    \label{fig:handling_techniques}
\end{figure}

Agile teams use various techniques to handle RCs as shown in Fig. \ref{fig:handling_techniques}. These include adding the RC to the product backlog, developing the RC as soon as received, following an acceptance criteria for RC, prioritization, re-estimating and changing the iteration plan. Additionally, we found that prioritizing based on changes in the market and based on team learning, and reviewing priorities every iteration as prioritization techniques. Furthermore, analysing the impact of the RC (step 1), estimating the effort of the RC (step 2), adding RC to iteration backlog (step 3), and removing existing user stories according to the priority and size of RC (step 4) as a step-wise technique of re-estimating and changing the iteration plan. We also found that, step 1 and 2 are used as acceptance criteria of RC.

\subsubsection{Managing the Agile Environment While Being Attentive of Requirements Changes}
Keeping in the mind the fact that RCs exist in the development environment, participants mentioned,
\begin{itemize}
\item Not over focusing on agile principles;
\item Planning iterations precisely;
\item Being agnostic of prescriptive agile frameworks;
\item Customizing and evolving roles and process
\end{itemize}
help them to manage the development environment better. The overall idea of the above mentioned indicate that, not trying to practice ideal agile help the teams to better handle the entire environment in terms of RCs.

%% file: sections/discussion.tex
\begin{figure*}
    \centering
    \includegraphics[width=\textwidth]{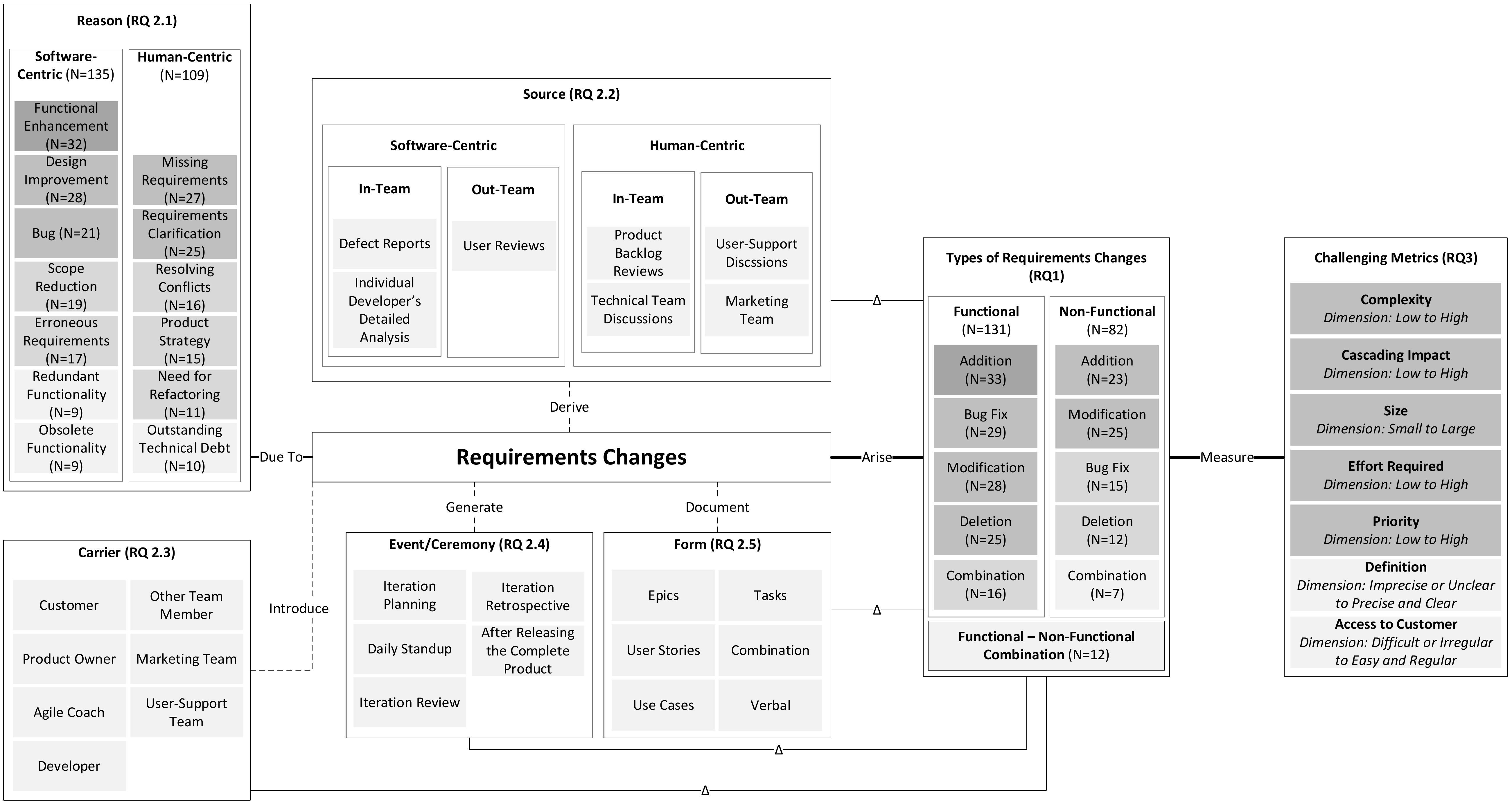}
    \caption{Conceptual Framework: Facets of Requirements Changes in Agile Environments ($\Delta$: Variate)}
    \label{fig:facets_framework}
\end{figure*}

Our findings provide insights to RCs as a journey within agile software development. As shown in Fig. \ref{fig:facets_framework}, reasons and sources give rise to RCs at different ceremonies/events in different forms and brought to the team by various carriers. And the teams are reluctant to receive RCs and tend to mitigate them with the use of different strategies (N=8). In case if they have to accept the RCs, they use several RC handling techniques (N=11) and also the teams face challenges when dealing with RCs. In addition, how challenging an RC is, can be identified with the use of the metrics we have given in this paper.

However, responding to changes is not the major reason for the teams to use agile in their work. Instead,  other reasons which eases the team in terms of work are the stimuli to practice agile, as reported to us. Key findings of our survey are given in Table \ref{tab:key findings}.

\begin{table}[]
\caption{Key Findings of the Study (RC: Requirements Change)}
\label{tab:key findings}
\resizebox{\columnwidth}{!}{%
\begin{tabular}{@{}lll@{}}
\toprule
\textbf{} & \multicolumn{1}{c}{\textbf{Key Finding}}                                                                                                                                                                                                                                                                       & \multicolumn{1}{c}{\textbf{Section}}    \\ \midrule
KF1       & \begin{tabular}[c]{@{}l@{}}Majority of RCs are functional RCs.\end{tabular} & \multicolumn{1}{c}{\ref{sec:type}}         \\ \midrule
KF2        & \begin{tabular}[c]{@{}l@{}}The types of RCs we provided the participants to select were \\ preliminary study and literature-based. Even though they had \\ the chance to provide any other classifications, everyone used\\ the given classification as is for RCs in agile except for bug fixes.\end{tabular} & \multicolumn{1}{c}{\ref{sec:type}}         \\ \midrule
KF3        & \begin{tabular}[c]{@{}l@{}}Majority of the reasons for RC origination are software-centric. \\ However, reasons such as inadequate communication, \\ documentation, rushed analysis, and wrong set of initial \\ requirements set use of agile in question.\end{tabular}                                       & \multicolumn{1}{c}{\ref{sec:reason}}       \\ \midrule 
KF4       & \begin{tabular}[c]{@{}l@{}}Apart from responding to changes, due to several other reasons \\ which are human-centric and software-centric, agile is practiced.\end{tabular}                                                                                                                                    & \multicolumn{1}{c}{\ref{sec:reason}}       \\ \midrule 
KF5        & \begin{tabular}[c]{@{}l@{}}Negatives such as stress due to iteration pressure, and not being\\ able to meet customer's expectations with the use of agile exist.\end{tabular}                                                                                                                                      & \multicolumn{1}{c}{\ref{sec:reason}}       \\ \midrule 
KF6        & \begin{tabular}[c]{@{}l@{}}Teams are reluctant to receive RCs and use techniques to mitigate\\ the origination of RCs.\end{tabular}                                                                                                                                                                            & \multicolumn{1}{c}{\ref{sec:reason}}       \\ \midrule 
KF7       & \begin{tabular}[c]{@{}l@{}}Acceptance criteria and techniques are used to handle RCs once \\ they are accepted.\end{tabular}                                                                                                                                                                                   & \multicolumn{1}{c}{\ref{sec:reason}}       \\ \midrule 
KF8        & \begin{tabular}[c]{@{}l@{}}By not over-focusing on agile principles, planning iterations \\ precisely, being agnostic of prescriptive agile frameworks, and \\ customizing and evolving roles and processes help to manage \\ agile environment in the presence of RCs.\end{tabular}                          & \multicolumn{1}{c}{\ref{sec:reason}}       \\ \midrule 
KF9        & Majority of sources of RCs are human-centric.                                                                                                                                                                                                                                                                  & \multicolumn{1}{c}{\ref{sec:source}}       \\ \midrule 
KF10        & \begin{tabular}[c]{@{}l@{}}RCs originate during daily standups most commonly than other \\ ceremonies/events.\end{tabular}                                                                                                                                                                                     & \multicolumn{1}{c}{\ref{sec:where}}        \\ \midrule 
KF11       & Customer plays the role of the carrier of the RC most of the time.                                                                                                                                                                                                                                             & \multicolumn{1}{c}{\ref{sec:who}}          \\ \midrule 
KF12       & \begin{tabular}[c]{@{}l@{}}Apart from agile focused forms such as epics and user stories, \\ use cases are also used to document RCs. Additionally, verbal \\ communication of RCs is also practiced to a very small extent.\end{tabular}                                                                      & \multicolumn{1}{c}{\ref{sec:form}}         \\ \midrule 
KF13       & \begin{tabular}[c]{@{}l@{}}Challenging nature of an RC can be measured through its \\ complexity, cascading impact, size, effort required, definition, \\ priority, and access to customer.\end{tabular}                                                                                                       & \multicolumn{1}{c}{\ref{sec:challenge}}    \\ \midrule
KF14       & Majority of the work load is not RCs.                                                                                                                                                                                                                                                                          & \multicolumn{1}{c}{\ref{sec:agile_helps}} \\ \bottomrule
\end{tabular}%
}
\end{table}

\textbf{Majority of RCs are functional RCs.} This resulted in 4 clues for the root cause.
\begin{itemize}
    \item All stakeholders are focused on functional requirements and/or;
    \item Non-functional requirements are implemented well and do not require to be changed and/or;
    \item Non-functional RCs are less common and/or;
    \item Less focus is given to non-functional requirements.
\end{itemize}

Taking both functional and non-functional RCs into consideration, the majority are additions. This suggests that stakeholders are interested in adding new requirements and leaves the question whether they are more interested in adding new requirements rather than improving the existing requirements.

Even though \textbf{bug fixes are not considered as a type of RC} as reported by some of the participants, bug fixes are prominent in agile environments according to our findings:

\begin{center}
    \textit{``I do not think bug fix can be a requirement change. Requirement is a user need/problem and design depicts the solution suggested to solve the problem. A bug fix is needed to correct a design not a requirement.'' - P12 [Business Analyst]}
\end{center}

If more and more bug fixes occur in every iteration, a team would not be able to deliver the product in a timely manner. Our study did not find the reasons for the origination of bug fixes in particular. However, since we found that time-boxed nature of agile stresses out team members, the time-boxed nature of agile could also be a reason for the rise of bug fixes. Knowing the root causes of bug fixes and remedying accordingly could help to minimize the amount of bug fixes.

\textbf{Majority of the sources of RC are human-centric.} i.e., RCs’ origination is not direct software-centric artefacts but activities where humans are heavily interacting with each other/collaboration, such as product backlog  reviews and technical team discussions. Therefore, as long as human interaction/collaboration is high, there is a chance of an RC to originate. This is shown where ``user reviews’’ which is an individual task where users provide the reviews individually, had less responses than collaborative tasks.

\textbf{Customer him/herself bring RCs to the team most of the time and product owner (agile team member who is responsible for the product backlog) brings RCs to the team on an average.} However, this varies from type of RC to RC. Some teams have the business analyst role yet present. Furthermore, other agile team members except for product owner and developer bring RCs to team was less than the average of the time. This shows that customer involvement with teams is at a high level in practice. This is expected in agile environments. However, Maierhofer et al.'s findings show that continuous customer integration does not mitigate negative effects of RC on project success in agile \cite{Maierhofer2010}. Additionally, business analyst not being the carrier in environments where the role is present, suggests that business analyst is not being able to play his/her role with customer collaboration, which might give a drop in his/her contentment of doing the job.

Even though there are several agile ceremonies, especially when iteration reviews are present for customers to provide feedback on deliverables, our findings show that \textbf{RCs are originated during daily standups most commonly than during other ceremonies/events.} As the customer is the leading carrier of RCs and the RCs originate more at daily standups, it suggests that customer is present at daily standups. It is surprising to know that a customer is present at daily standups if the agile team is not onsite. However, since we did not capture this information, it is unable to confirm whether this changes from the environment. Therefore, despite the environment, it is interesting to see customers being present at daily standups. This may however lead to several issues such as:
\begin{itemize}
    \item Team being uncomfortable in front of the customer (varies according to the experience, empowerment, and other human aspects such as team culture and personality)
    \item Team being emotionally reactive to what they agree/do not agree with customer’s requests
    \item A possible chance of a lowered productivity in the team during the day as an intervention of an RC is provided at the beginning of the day as it is natural to expect the new information to come to the mind from time to time. This might unease the team and may lead to a lower level of productivity
    \item Increased amount of time spent on daily standups. Daily standups are meant to be within 5-10 minutes and it is impractical to stay within a short time period when RCs are provided at the event
    \item The purpose of daily standup is to discuss what was done yesterday, what went wrong, and what is planned to do today. If the purpose of iteration reviews barters with the purpose of daily standups, the team environment may deviate from an agile environment to a chaotic environment which may lead to an inferior working environment
\end{itemize}

While epics, user stories, use cases, tasks, and combination of these forms are being used to document RCs, RCs in agile are most documented in the form of user stories and tasks. This partially agrees with Wang et al.'s findings on user stories and use cases being the mostly used form of requirements representation in agile \cite{Wang2014a}. Moreover, some responses were found for \textbf{verbal communication} as well. In such cases, depending on from whom to whom the new RC is communicated, a variation of the way of handling is expected. This links to the human aspects such as emotions and personality as well. In a hypothetical incident of where customer provides the RC to a team member who shows negative emotions mostly and/or has a negative personality, may not even communicate the new RC with the product owner and the team. In such cases, issues may occur as this is direct customer dissatisfaction. Also, in the case of a highly empowered developer receiving the RC from the customer but does not respect the agile practices may even directly develop the new RC without informing the team. In both these cases, a major impact on product backlog and the project is made on its scope, time, and budget.

While a majority of sources of RCs were human-centric, \textbf{majority of reasons for origination of RCs were software-centric.} This suggests that stakeholders are interested in enhancing the functionalities of the software more than any other reason. Even though the majority of the reasons are software-centric, human-centric issues such as missing requirements, requirements clarification, and conflict resolvent also exist in a considerable amount. As it is the customer who brings the RCs to the team most according to our findings, these human-centric issues as reasons depict that the level of engagement between customer and the team is probably low even though customers' attempt to involve in the process is high. This impacts the project as better commitment from stakeholders is required to decrease project failure \cite{Hoff2008}.

Similarly, reasons such as inadequate communication whereas intensive communication with customer lead to capture the requirements changes \cite{Wang2014a}, inadequate documentation, rushed analysis when defining requirements, and wrong set of initial requirements were also mentioned as reasons for requirements changes origination. Cao and Ramesh \cite{Cao2008} found that rapidly changing competitive threats, stakeholder preferences, software technology, and time-to-market pressures (which can be categorized as human-centric reasons) as the key reasons for rendering pre-specified requirements inappropriate, which we can name as the the root causes for the wrong set of initial requirements. As a remedy to this, Gall et al. \cite{Gall2006TowardsElicitation} proposed a framework in which the requirements elicitation meetings are recorded, important stakeholder statements are automatically extracted and stored in a database. However, their framework is suitable for traditional software development where specific events are set to elicit requirements. If their approach is used in agile, all ceremonies/events are required to be recorded, especially daily standups as our findings suggest that RCs are originated at daily standups often. In case, the size of the requirements database require attention as standups are done daily.

Our findings show that the \textbf{challenging nature of an RC can be measured by its complexity, cascading impact, size, effort required, priority, definition, and access to the customer.} These can be positive or negative depending on the situation. However, in an agile environment an RC is challenging when complexity is high, cascading impact is high, size is large, effort required is high, priority is high, definition is imprecise or unclear, and access to customer is difficult or irregular.

\textbf{Even though agile manifesto states the principle ``accepting changing requirements even late in development’’, agile teams tend to mitigate the origination of requirements changes.} The strategies they use are clear definition of requirements, proper use of tools, and practices according to our survey results. Not only agile practices, but also non-agile practices are being used to mitigate the RCs from occurring. This suggests that, agile teams are reluctant to receive RCs. 

This is confirmed by our results where participants mentioned that the reasons for practicing agile is not always about embracing changes and delivering fast, but agile being easy for the team to follow and helps in easing and prioritizing the work. However, \textbf{in the presence of RCs agile teams use RC handling techniques} as we found. This includes prioritization, which is based on changes in the market, team learning, and reviewing priorities after every iteration. Moreover, according to Hoff et al. \cite{Hoff2008}, fixing errors and cost benefit trade-offs are the key decision factors when prioritizing requirements which can be extend to RCs as well. However, according to Baruah \cite{Baruah2015}, there is no structured approach to manage RCs in agile.


\subsection{Threats to Validity}
\label{sec:threats}
\input{sections/threats}

%% file: sections/threats.tex



\textbf{External Validity:}
Equal distribution of the participants across the world did not take place in this study. As given in Section \ref{sec:Mtd}, most participants (65\%) represented Asia. Therefore, the results are biased towards one part of the world. This territorial bias hinders generalizing the findings to the entire global software development community, although such idealistic generalizations are hardly achievable in practice.

\textbf{Internal Validity:}
The drop rate of the survey completion was considerably high as 106 participants have started and only 42 have completed the survey fully. The main reason we found was that the survey was overly lengthy and it contained questions with matrices which took excessive amount of time. This was not reported by any of the participants who participated in the pilot study. However, when designing the survey, we placed the participant demographic questions at the end of the survey with the intention of giving participants enough time to answer the survey questions. But still the drop rate was considerably high. A positive consequence of this was we were able to gather in-depth information about RCs.

Even though we provided open-ended questions for the participants to provide other opinions they have apart from the answers given in the close-ended questions, we found that many have not taken the opportunity to give their opinions in the open-ended questions. Therefore, we deduce that collecting data through interviews enriches the findings.

\textbf{Construct Validity:}
Interview transcripts of the preliminary study, and answers to the open-ended questions were coded and analysed only by the first author. However, to mitigate the bias, at the end of each round of analysis, we discussed the emerged categories among three of us.

%% file: sections/recommendations.tex
Based on our findings, following are some recommendations for agile practitioners to follow when to when dealing with RCs.


     
     \textbf{Find root causes of bugs early:} Irrespective of bug fixes being/being not a type of RC (Section \ref{sec:type}), if you get bug fixes often, find the root cause and remedy the root cause. If the root cause is related to how agile is practiced, be flexible enough to change the environment accordingly.
    
    
    \textbf{Expect RCs to often come directly from customer:}
    Our findings suggest that, irrespective of customer being onsite or not, the customer reaches the agile team directly and provides RCs. Therefore, expect RCs directly from the customer rather than through product owner or some other stakeholder who interacts with customer often. This is especially important for teams who are transitioning from traditional software development to agile software development, and who practice hybrid software development (traditional and agile software development together).
    
    \textbf{Expect RCs at daily standups often:} Even though there are other agile ceremonies/events to get customer feedback, customers tend to provide the RCs mostly at the daily standup. If the rest of the process is not genuine, we recommend following another.
    
    \textbf{Accept rather than resist RCs:} RCs may direct the project in the right direction. Therefore, rather than trying to mitigate the RCs from originating, accept the appropriate RCs and use RC handling techniques to manage it effectively. We recommend analysing the impact of RC, estimate the effort required to develop it, prioritise, and discuss with the team and customer before accepting it (Section \ref{sec:agile_helps}). \textbf{Spending more time in requirements analysis will help to avoid wrong requirements and thus RCs.}
    
     \textbf{Understand the complexity of each RC:} We recommend to use the metrics: complexity, cascading impact, size, effort required, priority, definition, and access to customer to measure the challenging nature of RCs and remedy accordingly to reduce the challenging nature of RCs (Section \ref{sec:challenge}) .
     
    \textbf{Better documentation of RCs:} Avoid relying on verbal communication alone as non-documentation often results in project failure in terms of schedule, cost, and quality (Section \ref{sec:form}). Non-documentation may also result in the team misinterpreting the RC. Therefore, documenting RCs at a sufficient level is recommended.
    
    \textbf{Maintain sustained and adequate communication through out the project:} Communication has been shown in many studies to be a key construct for better software development~\cite{Gizzatullina2019, Abdullah2011CommunicationEngineering}. Inadequate communication commonly leads to the origination of unexpected RCs or more challenging RCs (Section \ref{sec:reason}), which then leads to potentially serious consequences affecting the cost, schedule, and quality of the project. In order to mitigate this, we recommend explicit efforts to maintaining adequate communication throughout the project.
    
    \textbf{Avoid introducing/accepting RCs in the middle of an iteration} as it causes interrupted team focus and impacts product delivery negatively (Section \ref{sec:agile_helps}). Even though agile does not recommend accepting RCs in the middle of an iteration, our findings show that this happens in practice. Therefore, we recommend not to accept or introduce RCs in the middle of an iteration, but to wait till the current iteration is completed to take necessary steps on deciding the acceptance of the RC.
    
    Even though agile is supposed to help responding to changes, in practice we recommend not to try practicing ``ideal agile" at the expense of the project outcomes (Section \ref{sec:agile_helps}). Be flexible around roles and processes in the agile environment. This is a better option than trying to make the agile environment an ideal or model agile environment for the sake of it. This will make for a more comfortable working environment than a stressful one trying to achieve some theoretic ideal of a  agile environment \cite{Masood2020RealPractice}.



\textbf{Researchers talking the language of agile practitioners:}
Our survey distribution approach included posting the survey link on social media and resistance to our survey was shown by a set of agile practitioners in a \textit{LinkedIn} group. This resulted in showing a gap between agile practitioners and agile researchers. The main gap we found was with respect to the terminology we used. Since our survey terminology contained the word ``requirement'', some of the practitioners were highly skeptical about using it as they preferred calling RCs ``user stories'' rather than ``requirements''. This is counter-intuitive as we considered user stories as a form of documenting requirements and RCs. A debate among some of the practitioners identified that there is no proper terminology set up to be used in agile environments and this hinders communication. It is necessary to look into this issue more in the future. Furthermore, practitioners seem to have different opinions on the definition of an RC. While some said all the requirements they receive are RCs, some others said there are no such entity as a requirements change in agile and they are all regular ``requirements". Another commented that all RCs turn out to be requirements in the end. Researchers need to understand that there is no proper definition for what agile teams receive as changes from customers or what we call requirements in general. Moreover, some practioners also showed their dislike of the concept of trying to manage RCs in agile as well.  

\textbf{Future Work:} Is agile as human-centric as it is known to be? As we found a participant mentioning that iteration pressure causes stress within team members, this requires further investigation majorly (Section \ref{sec:agile_helps}).



%% file: sections/relatedwork.tex







McGee and Greer's studies \cite{McGee2011SoftwareStudy,McGee2009ATaxonomy} resulted in an RC taxonomy for software development based on sources categorized as RCs originating from \textit{market, organization, project vision, specification,} and \textit{solution}. Their study found that higher cost and value changes originate mostly from \textit{organization} and \textit{vision} \textit{sources} and these sources involve cooperation of stakeholder groups with less control than the RCs originating from the sources: \textit{specification} and \textit{solution}. Additionally, except for RCs originating from market, RCs originating from other sources showed a considerable amount of difference in cost, value to the customer, and management considerations. They also found the triggers and uncertainties for each classified source. Furthermore, findings of these studies indicate that using the given taxonomy will help to manage RCs, understand RCs, and gain risk visibility.

Nurmuliani et al.'s work \cite{Nurmuliani2004AnalysisCycle}, presents types of RCs, reasons of they originate, and sources from which RCs originate as its key findings. They categorized types of RCs as additions, deletions, and modifications of requirements. Also, they found that defect fixing, missing requirements, functionality enhancement, product strategy, design improvement, scope reduction, redundant functionality, obsolete functionality, erroneous requirements, resolving conflicts, and clarifying requirements as the reasons of why RCs originate. As sources of RCs, they found defect reports, engineering's calls, project management consideration, marketing group, developers' detailed analysis, design review feedback, technical team discussion, functional specification review, feature proposal review, and  customer-support discussions. In addition to the taxonomy based on type, reasons, and sources of RC, they also provided change request arrival rate, and requirements volatility measure which is the ratio of number of RC to the total number of RC over a certain period of time (during development phase).

Harker et al. found mutable, emergent, consequential, adaptive, and migration as RCs in their classification \cite{Harker1993TheEngineering}. They also found environmental turbulence, stakeholder engagement in requirements elicitation, system and user development, situation action and task variation, and constraints of planned organizational development as the \textit{origins} of RCs.

Inpirom and Prompoon classified RCs according to analysis and design of software artefacts \cite{Inpirom2013DiagramUML}. Their taxonomy is based on unified modeling language including the most commonly used diagrams: use case diagram, class diagram, and sequence diagram. In addition, they mentioned that most research focused on impact of RCs on source code but impact of RCs on diagrams are also required. Likewise, Basirati et al.'s study \cite{Basirati2015UnderstandingStudy} indicate that, RCs impact artefacts. Therefore, they explored the changes in use cases and their further analysis of problematic changes resulted in a taxonomy of RC. In addition, they also found the local and temporal dispersion of RC as difficult and risky.

The only taxonomy we found for RCs in agile software development was Saher et al.'s literature based study \cite{Saher2018}. They described their taxonomy in terms of \textit{time of change, type, reason,} and \textit{origin} of RC. However, out of the 9 studies they used to build the taxonomy, 8 use traditional software development  while only one study actually focused on agile methods. This leaves a gap in the research where a taxonomy of RC in agile is necessary where responding to RCs is paramount and in this paper we have filled that gap.


\begin{table*}[t]
\caption{Comparison of Our Study with Related Work}
\label{tab:gap}
\centering
\resizebox{1.5\columnwidth}{!}{%
\begin{tabular}{@{}lllllllll@{}}
\toprule
\textbf{}            & \textbf{Type} & \textbf{Reason} & \textbf{Source} & \textbf{\begin{tabular}[c]{@{}l@{}}Ceremony/\\ Event\end{tabular}} & \textbf{Carrier} & \textbf{Form} & \textbf{\begin{tabular}[c]{@{}l@{}}Impact on \\ Artefacts\end{tabular}} & \textbf{\begin{tabular}[c]{@{}l@{}}Challenging \\ Metrics\end{tabular}} \\ \midrule
\multicolumn{9}{l}{\textbf{Requirements Changes Taxonomies in Traditional Software Development:}}                                                                                                                                                                                                                                    \\ \midrule
McGee and Greer \cite{McGee2009ATaxonomy, McGee2011SoftwareStudy}     &               &                 & \checkmark          &                                                                    &                  &               &                                                                         &                                                                         \\
Nurmuliani et al. \cite{Nurmuliani2004AnalysisCycle}   & \checkmark             & \checkmark              & \checkmark              &                                                                    &                  &               &                                                                         &                                                                         \\
Harker et al. \cite{Harker1993TheEngineering}       & \checkmark             &                 &                 &                                                                    &                &               &                                                                         &                                                                         \\
Iniprom and Prompoon \cite{Inpirom2013DiagramUML} &               &                 &                 &                                                                    &                  &               & \checkmark                                                                       &                                                                         \\ \midrule
\multicolumn{9}{l}{\textbf{Requirements Changes Taxonomies in Agile Software Development:}}                                                                                                                                                                                                                                          \\ \midrule
Saher et al. \cite{Saher2018}        & \checkmark             &  \checkmark             & \checkmark               &                                                                    &                  &               &                                                                         &                                                                         \\
\rowcolor[HTML]{EFEFEF} 
\textbf{Our study}   & \checkmark    & \checkmark      & \checkmark      & \checkmark                                                         & \checkmark       & \checkmark    & \textbf{}                                                               & \checkmark                                                              \\ \bottomrule
\end{tabular}%
}
\end{table*}

%% file: sections/conclusion.tex

In this paper we have presented a conceptual framework of requirements changes in agile to provide a holistic picture of requirements changes in agile software development environments. Through an interview-based preliminary study with participation of 10 agile practitioners in New Zealand and Australia, existing literature, and an in-depth survey with the participation of 40 agile practitioners across the globe, we found the types of requirements changes, reasons, sources, carriers, forms of requirements changes, metrics to figure out the challenging nature of agile, and whether agile helps to respond to requirements changes or not. Furthermore, we found that bug fixes are prominent in agile environments. Cross-functionality, organizational needs, and time-boxed nature of agile intensifies the challenging nature of requirements changes. Also, a majority of the sources of requirements changes are human-centric but a majority of the reasons for leading to requirements changes are software-centric. However, reasons such as inadequate communication, documentation, rushed analysis, and wrong of initial requirements question the proper use of agile. In addition, we found that requirements changes originate during daily standups most commonly than during other ceremonies/events. Also, the customer brings requirements changes to the team as compared to any other stakeholder. Furthermore, we found that responding to change is not the major reason for teams to use agile in their software development and they use agile as it eases their work. Ultimately, even though most teams provide less focus to the agile value ``responding to change over following a plan'' and the principle ``accepting changing requirements even late in development''; not over focusing on agile principles, planning iterations precisely, being agnostic of prescriptive agile frameworks, customizing and evolving roles and processes will help to manage the agile environment in the presence of requirements changes.